\documentclass[preprint,3p]{elsarticle}
\pdfoutput=1

\usepackage{graphicx}
\usepackage{amssymb, amsmath, amsxtra}

\usepackage{epsfig,subfigure}

\usepackage[T2A]{fontenc}
\usepackage[utf8]{inputenc}
\usepackage[english]{babel}

\usepackage{graphicx}
\usepackage{enumitem}

\usepackage{lscape}
\usepackage{pdflscape}

\usepackage{hyperref}

\def\dh#1{\mathop {#1}\limits_{h}}

\def\dsm#1{ \mathop{#1}\limits_{-s}}

\def\dh#1{ \mathop{#1}\limits_ h}
\def\ds#1{\mathop {#1}\limits_{-s}}
\def\dsp#1{\mathop {#1}\limits_{+s}}
\def\dtau#1{\mathop {#1}\limits_{-\tau}}
\def\dtaup#1{\mathop {#1}\limits_{+\tau}}

\def\dh#1{\mathop {#1}\limits_h}
\def\dtau#1{\mathop {#1}\limits_{-\tau}}
\def\dtaup#1{\mathop {#1}\limits_{+\tau}}

\def\dvhb#1{\dh#1_{\bar 1}}
\def\dvhb2#1{\dh#1_{\bar 2}}

\def\dh#1{\mathop {#1}\limits_{h}}
\def\dtau#1{\mathop {#1}\limits_{-\tau}}
\def\dtaup#1{\mathop {#1}\limits_{+ \tau}}

\def\dvhb#1{\dh#1_{\bar x}}

\def\dtau#1{\mathop {#1}\limits_{-\tau}}
\def\dtaup#1{\mathop {#1}\limits_{+ \tau}}

\def\dh#1{\mathop {{#1}}\limits_{h}}

\def\dh#1{ \mathop{#1}\limits_ h}

\def\DSP{\dsp{D}}
\def\DSM{\ds{D}}

\def\DTP{\dtaup{D}}
\def\DTM{\dtau{D}}

\def\SSP{\dsp{S}}
\def\SSM{\dsm{S}}
\def\STP{\dtaup{S}}
\def\STM{\dtau{S}}

\newcommand{\ppartial}[1]{\frac{\partial}{\partial #1}}

\newtheorem{remark}{Remark}

\begin{document}

\begin{frontmatter}

\title{Shallow water equations in Lagrangian coordinates: symmetries,
conservation laws and its preservation in difference models}

\author[mymainaddress]{V.~A. Dorodnitsyn\corref{mycorrespondingauthor}}
\cortext[mycorrespondingauthor]{Corresponding author}
\ead{Dorodnitsyn@keldysh.ru,dorod2007@gmail.com}

\author[mymainaddress,mysecondaryaddress]{E.~I. Kaptsov}
\ead{evgkaptsov@gmail.com}

\address[mymainaddress]{Keldysh Institute of Applied Mathematics,\\
Russian Academy of Science, Miusskaya Pl. 4, Moscow, 125047, Russia}

\address[mysecondaryaddress]{School of Mathematics, Institute of Science, \\
Suranaree University of Technology, 30000, Thailand}

 \begin{abstract}
%Lie point symmetries of shallow water equations in Eulerian and
%Lagrangian coordinate systems are considered. The Noether theorem
%and direct method yield conservation laws which can be preserved in
%the invariant difference schemes and meshes.
%
%\medskip

The one-dimensional shallow water equations in Eulerian and Lagrangian coordinates are considered.
It is shown the relationship between symmetries and conservation laws in Lagrangian (potential)
coordinates and symmetries and conservation laws in mass Lagrangian variables.
For equations in Lagrangian coordinates with a flat bottom an invariant difference scheme is constructed
which possesses all the difference analogues of the conservation laws: mass, momentum, energy, the law of
center of mass motion.
Some exact invariant solutions are constructed for the invariant scheme,
while the scheme admits reduction on subgroups as well as the original system of equations.
For an arbitrary shape of bottom
it is possible to construct an invariant scheme with conservation of mass and momentum or energy.
Invariant conservative difference scheme for the case of a flat bottom tested numerically in comparison
with other known schemes.
\end{abstract}


\end{frontmatter}

\bigskip

%\cite{DORODNITSYN2019201,KOZLOV2019,bk:Korobitsyn_scheme[1989]}

\section{Introduction}

The shallow water equations describe the motion of incompressible fluid
in the gravitational field if the depth of the liquid layer is small enough.
They are widely used in the description of processes in the atmosphere,
water basins, modeling of tidal oscillations, tsunami waves and gravitational
waves~(see the classical papers such as~\cite{bk:Whitham[1974],bk:Ovsyannikov[2003]} and
detailed description in, for example,~\cite{bk:PetrosyanBook[2010],bk:Vallis[2006]}).
Even in case of the one-dimensional shallow water equations with the flat bottom
one meets certain difficulties  to obtain nontrivial exact solutions.
Some exact solutions can be found in~\cite{bk:PetrosyanBook[2010],bk:Bernetti[2008],bk:HanHantke[2012]}.

Therefore, numerical calculations and finite-difference modeling are an effective mathematical
apparatus in that area. Some monographs~\cite{bk:KulikovskyPogorelovSemenov,bk:Vreugdenhil[1994],bk:TanWeiyan,bk:Abbasov}
and numerous articles, e.~g.,~\cite{bk:YeleninKrylov[1982],bk:Bihlo_numeric[2012],
bk:Bihlo_numeric[2017],bk:Bihlo_numeric[2019],
bk:MurshedFutai[2019],bk:KhakimzyanovIV,
bk:DyakonovaKhoperskov,bk:MoralesCastro},
are devoted to the numerical modeling of processes described by shallow water equations.

The present paper is devoted to the construction of invariant conservative difference schemes
for the shallow water equations in potential and mass Lagrangian coordinates.

Lie groups have provided efficient tools for studying ODEs and PDEs since the fundamental works
of Sophus Lie~\cite{Lie15, bk:Lie[1891b], bk:Lie-Scheffers[1896]}. The symmetry group of a differential
equation transforms solutions into solutions while leaving the set
of all solutions invariant. The symmetry group can be used to obtain
new solutions from known ones and to classify equations into
equivalence classes according to their symmetry groups. It can also
be used to obtain exact analytic solutions that are invariant under
some subgroup of the symmetry group, so called ``group invariant
solutions''. Most solutions of the nonlinear differential equations
occurring in physics, mechanics and in other applications were obtained in this
manner. Applications of Lie group theory to differential equations,
known as group analysis, is the subject of many books and review
articles~\cite{bk:Ovsyannikov[1962],bk:Olver,bk:Ibragimov1985,bk:Bluman1989,
bk:HandbookLie_v1,bk:Gaeta1994}.

The group properties of the shallow water equations were studied in numerous papers (see~\cite{bk:HandbookLie_v2,bk:LeviNicciRogersWint[1989],bk:ClarksonBila[2006]}).
Group classification and first integrals of these equations can
be found in~\cite{bk:AksenovDruzkov_classif[2019],bk:KaptsovMeleshko_1D_classf[2018]}.
It was shown~(see, e.~g.,~\cite{bk:KaptsovMeleshko_1D_classf[2018]}) that the shallow
water equations in Lagrangian coordinates can be obtained
as Euler--Lagrange equations of Lagrangian functions of a special kind.
Extended nonlinear models, such as the Green--Naghdi equations and modified shallow water equations
from a group point of view were considered in~\cite{bk:SiriwatKaewmaneeMeleshko2016, bk:SzatmariBihlo[2014]}.

More recently applications of Lie groups have been extended
to difference equations~\cite{Maeda1, Maeda2,Dor_1, Dor_2, Dor_3,
bk:DorodKozlovWint[2004],[LW-2],bk:DorodKozlovWinternitz[2000], Quisp, [LW-3], bk:Dorodnitsyn[2011],
Vinet, bk:Hydon_book[2014],bk:DorodKozlovWintKaptsov[2015]}. The applications are quite the same as in the case of differential equations,
but have certain peculiarities connected with nonlocal character of difference operators and
geometrical structure of difference mesh~\cite{bk:Dorodnitsyn[2011]}.

 The invariant finite-difference schemes that preserve sym\-met\-ries and since the
geometric properties of the original equations  are of  particular interest.
For the shallow water equations such schemes on moving
meshes were proposed in~\cite{bk:Bihlo_numeric[2012]}~(for the methods of introducing moving meshes,
see~\cite{bk:BuddHuang[1996],[Huang]}).
The difficulties  of constructing energy-saving difference schemes for the
shallow water equations was emphasized in~\cite{bk:Bihlo_numeric[2012]},
and there the references were made to non-invariant schemes possessing the conservation
law of energy~(see, for example,~\cite{bk:ArakawaLamb[1981]}).

The Lagrange mass coordinates\footnote{To avoid terminological confusion,
further we also refer standard Lagrangian coordinates as~\emph{potential} coordinates
to distinguish from the mass Lagrangian coordinates.}
~\cite{bk:SamarskyPopov_book[1992],bk:YanenkRojd[1968]}
are widely  used in numerical modeling of the equations of continuous medium.
Notice that Lagrangian mass coordinates are related with standard  Lagrangian coordinates by tangent transformations
and since may not preserve the whole set of admitted Lie group of point transformations.
In the papers~\cite{bk:SamarskyPopov[1969],bk:SamarskyPopov_book[1992]}
completely conservative schemes for the gas dynamics equations in Lagrangian
mass coordinates were constructed.
One can find examples of such schemes in gas dynamics and magnetohydrodynamics in~\cite{bk:SamarskyPopov[1969],KolPovPop87,bk:Poveschenko[2019],bk:PopovPoveschenkoPolyakov[2017],
DORODNITSYN2019201,KOZLOV2019,bk:Korobitsyn_scheme[1989]}.
%Completely conservative equation scheme for \emph{bilayer} shallow water in Lagrange coordinates
%was built in \cite{bk:YeleninKrylov[1982]}.
\medskip

The paper is organized as follows.
In the second section basic equations in Eulerian and Lagrangian mass coordinates
and some introductory remarks are given.
The third section discusses in detail the shallow water
equations in Lagrangian~(potential) coordinates
and their connection to the Lagrangian mass coordinates.
In the third section, invariant difference schemes
possessing local conservation laws of energy, mass, center of mass
and momentum for the shallow water equations with the flat bottom are constructed.
For the case of an arbitrary bottom, difference schemes possessing
local conservation laws of mass and energy~(or momentum) are presented.
For the flat bottom equations some invariant difference solutions
and corresponding reductions of the scheme are given.
In the fifth section, one of the constructed invariant schemes is numerically
 performed on several test problems. The same
tests are performed on some know schemes, and the obtained results
are compared.
In Conclusion the obtained schemes and numerical results are discussed.

\section{Shallow water equation for Eulerian and Lagrangian mass coordinates}

System of the one-dimensional shallow water equations with an arbitrary bottom
in Eulerian coordinates has the following form
\begin{equation} \label{Euler1}
\eta_t + ((\eta +h)u)_x = 0,
\end{equation}
\begin{equation} \label{Euler2}
u_t +uu_x + \eta_x=0.
\end{equation}
where $u(t,x)$ is the velocity of the continuous medium particles,
$\eta(t,x)$ is the height of a liquid column above the bottom at point $x$,
and the bottom profile is described by the function~$h(x)$.
One can reduce the linear bottom case~($h(x) = k x$, where $k$ is constant)
to the flat bottom~($h = 0$) by the following change
of variables~\cite{bk:ChirkunovPikmullina[2014]}
\[
    \tau = k t,
    \quad
    \xi = k \left(x - \frac{k t^2}{2}\right),
    \quad
    u = \nu(\tau,\xi)+\tau,
    \quad
    \eta = w(\tau,\xi) - \xi - \frac{\tau^2}{2}
\]
In case of the flat bottom, it turns out that
system~(\ref{Euler1}),(\ref{Euler2}) possesses especially simple form
with the help of hodograph transformation
\large
\begin{equation}\label{ek:hodograph_Euler}
\def\arraystretch{1.5}
\begin{array}{c}
    x = x(\rho, u),
    \qquad
    x_\rho = -\frac{u_t}{\Delta},
    \qquad
    x_u = \frac{\rho_t}{\Delta},
    \\
    t = t(\rho, u),
    \qquad
    t_\rho = \frac{u_x}{\Delta},
    \qquad
    t_u = -\frac{\rho_x}{\Delta},
    \\
    \Delta = \rho_t u_x - \rho_x u_t \neq 0
\end{array}
\end{equation}
\normalsize
that allows one to linearize~\cite{bk:YanenkRojd[1968],bk:Ovsyannikov[2003]}
the system into the following equations
\[
    x_u - u t_u + {\rho} t_{\rho} = 0,
\]
\[
    x_{\rho} + t_u - u t_{\rho} = 0.
\]
%\subsection{Eulerian coordinates}
%Shallow water equations in Euler's coordinates read
%\begin{equation} \label{Euler1}
%\eta_t + ((\eta +h)u)_x = 0,
%\end{equation}
%\begin{equation} \label{Euler2}
%u_t +uu_x + \eta_x=0,
%\end{equation}
%where $u(t,x)$ and $\eta(t,x)$ are dependent variables of $t,x$,
%$h=h(x)$ is a given function of $x$ which describes a shape of
%bottom.

\bigskip

Now we involve the Lagrangian operator of total differentiation with
respect to $t$
\begin{equation} \label{dif}
\frac{d}{dt}= D_t + u D_x,
\end{equation}
which does not commute with $D_x$:
$$
\left[\frac{d}{dt}, D_x\right] \neq 0.
$$
Along with $\frac{d}{dt}$ we introduce two variables: a ``density''
$\rho$
\begin{equation} \label{var}
\rho = h+\eta,
\end{equation}
and a new independent (mass) coordinate $s$ by means of {\it
contact} transformation
\begin{equation} \label{s}
ds = \rho dx - \rho u dt,
\end{equation}
where $ds$ is a total differential form, i.e.
\begin{equation} \label{tot}
\frac{\partial \rho}{\partial t} =    - \frac{\partial(\rho
u)}{\partial x},
\end{equation}
that equivalent to the equation
\begin{equation} \label{cont}
\rho_t + (\rho u)_x =0.
\end{equation}
Collecting all relations we get the following transform:
$$
\frac{dx}{dt}=u, \quad  \frac{\partial x}{\partial s}=
\frac{1}{\rho}, \quad  \frac{\partial s}{\partial t}=-\rho u.
$$

We introduce the following operator of total differentiation with
respect to $s$:
\begin{equation} \label{D_s}
D_s = \frac{1}{\rho}D_x.
\end{equation}
The operators $\frac{d}{dt},D_s$ commute on the system
(\ref{Euler1}),(\ref{Euler2}):
\begin{equation} \label{oper}
\left[\frac{d}{dt}, D_s\right]
= \left[D_t+ u D_x, \frac{1}{\rho}D_x \right]
= -\frac{1}{\rho^2}\left(\rho_t + (\rho u)_x\right) ,
\end{equation}
where the last equation is zero on equation~(\ref{Euler1}).

So far we can change variables as following:
\begin{equation} \label{change}
\displaystyle
\def\arraystretch{1.5}
\begin{array}{c}
    u(t,x)=u(t,s),
    \quad
    \rho(t,s) = h(x)+\eta(t,x),
    \\
    \frac{du}{dt} = u_t + uu_x,
    \quad
    \frac{d\rho}{dt} = \eta_t + u(h+\eta)_x,
    \\
    u_x=\rho u_s,
    \quad
    \rho_x=\rho \rho_s,
    \quad
    ds = \rho dx - \rho u dt,
    \quad
    t=t.
\end{array}
\end{equation}
 Now we can rewrite the original system in the Lagrangian
coordinates $(t,s,u,\rho)$:
\begin{equation} \label{Lag1}
 \frac{d}{dt}\left(\frac {1}{\rho} \right) - u_s =0,
\end{equation}
\begin{equation} \label{Lag2}
 \frac{du}{dt} + \rho(\rho-h)_s =0.
\end{equation}
%Notice, that the last system possesses an additional symmetry
%\begin{equation} \label{symmetry}
%X_1=\frac{\partial}{\partial t},\quad X_2=\frac{\partial}{\partial
% s}, \quad X_3=\frac{\partial}{\partial u}.
%\end{equation}

\section{Shallow water equation for (potential) Lagrangian coordinates}

By means of a contact transformation one can relate Lagrangian mass coordinates
to potential Lagrangian coordinates which are of our main interest. Potential coordinates allow
one to derive the shallow water equations as Euler-Lagrange equations
of a specific Lagrangian function. What is more important, they preserve orthogonality
of a difference mesh for all the symmetries of the shallow water equations that
significantly simplifies the subsequent discretization procedure.

\subsection{Shallow water equation with an arbitrary bottom profile}
\label{sec:sw_arb_bottom}

We consider a potential $x$, defined by
\begin{equation} \label{pot}
\frac{dx}{dt} =u; \quad \frac{\partial x}{\partial s}=
\frac{1}{\rho}.
\end{equation}
Then continuity equation (\ref{Lag1}) reads
\begin{equation}\label{x_commute}
  x_{ts}= x_{st},
\end{equation}
while equation~(\ref{Lag2}) will be the following (everywhere below the
derivative with respect to $t$ is a Lagrangian one):
\begin{equation} \label{potents}
x_{tt} - \frac{x_{ss}}{x_s^3} - h_x=0.
\end{equation}

\begin{remark}
Notice that equation~(\ref{potents}) corresponds to one-dimensional gas
dynamics of polytrophic gas with $\gamma = 2$.
Gas dynamics equations in this case are augmented with a state equation of the form
\[
  p = A \rho^\gamma,
\]
where $p$ is the pressure and $A$ is constant. For the hyperbolic shallow
water  equations one can put following ``state equation'':
\cite{Ovsiannikov[2003]}
\begin{equation}\label{ek:sw_stetEq}
  p = \frac{1}{2} \rho^2,
\end{equation}
which allows one to rewrite system (\ref{Lag1}), (\ref{Lag2})
in an alternative form:
\[
 \frac{d}{dt}\left(\frac {1}{\rho} \right) - u_s =0,
    \qquad
 \frac{du}{dt} + p_s = h_x.
\]

 %Equation (\ref{ek:sw_stetEq})
%is especially useful in numerical schemes construction.
\end{remark}

The  equation (\ref{potents}) admits two symmetries
\begin{equation}\label{potents_syms}
  X_1 = \frac{\partial}{\partial t}
  \quad
  \mbox{and}
  \quad
  X_2 = \frac{\partial}{\partial s}.
\end{equation}
Equation (\ref{potents})
can be considered as Euler-Lagrange equation for the Lagrangian
\begin{equation} \label{potents1}
{\cal L}= \frac{x_t^2}{2}-\frac{1}{2x_s}+h(x).
\end{equation}

Now the Noether theorem can be applied. First, we consider the case
where~$h=h(x)$ is an {\it arbitrary} function.
In this case one obtains the following conservation laws:

\begin{enumerate}
\item
$X_1=\frac{\partial}{\partial t}:$
\begin{equation} \label{CL1}
\frac{d}{dt}\left[ \frac{1}{2x_s}+\frac{x_t^2}{2}-h(x)\right]+  D_s\left[
\frac{x_t}{2x_s^2}\right]=x_t\{x_{tt} - \frac{x_{ss}}{x_s^3} - h_x\}=0;
\end{equation}

\item $X_2=\frac{\partial}{\partial s}:$
\begin{equation} \label{CL2}
\frac{d}{dt}\left[ x_t x_s\right]+  D_s\left[\frac{1}{x_s}-\frac{x_t^2}{2}-h(x)\right]=x_s\{x_{tt}
- \frac{x_{ss}}{x_s^3} - h_x\}=0.
\end{equation}

\end{enumerate}

Now we recalculate the above conservation laws of the system
(\ref{Lag1}), (\ref{Lag2}) in Lagrangian mass coordinates.

\begin{enumerate}
\item $X_1=\frac{\partial}{\partial t}:$
\begin{equation} \label{CLaw1}
\frac{d}{dt}\left[ \frac{u^2+\rho}{2}-h(x)\right]+  D_s\left[ \frac{\rho^2 u}{2}\right]=0.
\end{equation}

\item $X_2=\frac{\partial}{\partial s}:$
\begin{equation} \label{CLaw2}
\frac{d}{dt}\left[ \frac{u}{\rho}\right]+ D_s\left[\rho -\frac{u^2}{2}-h(x)\right] =0.
\end{equation}

\end{enumerate}

\begin{remark}
In verification of above conservation laws one should
keep in mind:
\begin{equation} \label{dx}
dh=h^\prime(x)dx = h^\prime \left(  u dt + \frac{ds}{\rho} \right),
\end{equation}
\end{remark}

\medskip

Let us recalculate the conservation laws
for the Euler coordinate system. For the
first one we have:
\begin{equation} \label{CLforE}
D_t^e\left[\rho \left( \frac{u^2+\rho}{2}-h(x) \right)\right]
+ D_x\left[\frac{\rho^2u}{2}+\rho u
\left(
    \frac{u^2 +     \rho}{2}-h(x)
\right)\right] =0.
\end{equation}
or
\begin{equation} \label{dxEuler}
\rho u (u_t +uu_x + (\rho-h)_x)+(\rho_t +(\rho
u)_x)\left(
    \rho + \frac{u^2}{2}-h
    \right)=0,
\end{equation}
where $D_t^e$ is the Euler total derivative with respect to $t$.

\medskip

The second conservation law in the Euler coordinate
system equals just the following
\begin{equation} \label{dx23}
D_t^e(u) +D_x\left(\rho -h +\frac{u^2}{2}\right)=0,
\end{equation}
or
$$
u_t+uu_x+\eta_x=0.
$$

\subsection{The case of a linear bottom}

Let us consider the linear bottom, i.e.,
\[
  h(x) = C_1 x + C_2,
\]
where $C_1$ and $C_2$ are constant.
Equation (\ref{Lag2}) becomes
\[
 \frac{du}{dt} + \rho(\rho - C_1 x)_s =0,
\]
or, in potential coordinates,
\[
  x_{tt} - \frac{x_{ss}}{x_s^3} - C_1 = 0.
\]
One can cancel  the constant $C_1$ by means of transformation
\begin{equation} \label{ek:lin_to_flat}
  x = \tilde{x} + \frac{t^2}{2} C_1,
\end{equation}
and arrive  to the case of the flat bottom~$h(x) = 0$
\begin{equation} \label{ek:elSW0}
  x_{tt} - \frac{x_{ss}}{x_s^3} = 0.
\end{equation}
For the sake of brevity, here and further on symbol ``\verb|~|'' is omitted.

Equation~(\ref{ek:elSW0}) admits the following symmetries
\begin{equation} \label{ek:syms6}
\def\arraystretch{1.5}
\begin{array}{c}
    X_1 = \ppartial{t},
    \quad
    X_2 = \ppartial{s},
    \quad
    X_3 = \ppartial{x},
    \quad
    X_4 = t\ppartial{x},
    \\
    X_5 = 3 t \ppartial{t} + 2 x\ppartial{x},
    \quad
    X_6 = 3 s \ppartial{s} + x \ppartial{x},
\end{array}
\end{equation}
and can be considered as Euler-Lagrange equation for the Lagrangian
\begin{equation} \label{ek:elL}
{\cal L}= \frac{x_t^2}{2}-\frac{1}{2x_s},
\end{equation}
which is invariant to the group actions of the symmetries $X_1$--$X_4$.
Notice, that (\ref{ek:elL}) is divergent--invariant to the action of the symmetry $X_4$, i.e.,
\[
  X_4 {\cal L} + {\cal L}\left(D_t\,\xi^t_4 + D_s\,\xi^s_4\right) = D_t(-x).
\]

Along with the Noether theorem there exists so called direct method~\cite{bk:BlumanAnco,bk:BlumanCheviakovAnco} which operates with the following relation:
\[
    D_t(F^t) + D_s(F^s) = \Lambda F,
\]
where $F$ is the original equation and $\Lambda$ is called an integrating multiplier.
As far an action of variational operator cancels any divergent expression,
then we get the determining equation for the multiplier $\Lambda$:
\[
    \textsf{E}(\Lambda F) \equiv 0,
\]
where the variational operator $\textsf{E}$ has the form
\[
\displaystyle
\textsf{E} = \ppartial{x} + \sum_{k=1}^\infty (-1)^k D_{i_1} \cdots D_{i_k} \ppartial{x_{i_1 \cdots i_k}}.
\]

Using Noether's theorem as well as the direct method  one can obtain the following conservation laws.
\begin{enumerate}
    \item
    For the symmetry $X_1 = \ppartial{t}$ on gets
    the conservation law of energy (\ref{CL1}), where $h(x) = 0$.

    The corresponding integrating factor is $\Lambda_1 = u_t$.
    %\begin{equation}
%      D_t\left(\frac{x_t^2}{2} + \frac{1}{2}{x_s^{-1}} \right)
%        + D_s\left(\frac{1}{2} x_t x_s^{-2}\right)
%        = x_t \left\{\frac{x_{ss}}{x_s^3} - x_{tt} \right\} = 0,
%    \end{equation}

    \item
    For $X_2 = \ppartial{s}$
    %\begin{equation}
%      D_t\left( x_t x_s \right)
%        + D_s\left( x_s^{-1} - \frac{1}{2} x_t^2 \right)
%        = x_s \left\{\frac{x_{ss}}{x_s^3} - x_{tt} \right\} = 0,
%    \end{equation}
    one obtains the conservation of mass (\ref{CL2}), where again $h(x) = 0$,
    and $\Lambda_2 = u_x$
    (it was shown in Section~\ref{sec:sw_arb_bottom}
    that the conservation law (\ref{CL2}) in Euler coordinates can be
    directly derived from the commutation relation $x_{ts}=x_{st}$).

    \item
    $X_3 = \ppartial{x}:$
    \begin{equation}\label{ek:CL3}
    \frac{d}{dt}(x_t) + D_s\left(\frac{1}{2}x_s^{-2}\right)
        = \frac{x_{ss}}{x_s^3} - x_{tt} = 0,
    \end{equation}
    with
    $\Lambda_3 = 1$,
    which is conservation of momentum.

    \item
    $X_4 = t\ppartial{x}:$
    \begin{equation}\label{ek:CL4}
    \frac{d}{dt}(t x_t - x) + D_s\left(\frac{1}{2} t x_s^{-2}\right)
        = t\left\{ \frac{x_{ss}}{x_s^3} - x_{tt}\right\} = 0,
    \end{equation}
    with
    $\Lambda_4 = t$,
    which means law of the center of mass motion.
\end{enumerate}

In Lagrangian mass coordinates the conservation
laws (\ref{ek:CL3}) and (\ref{ek:CL4}) have the following forms
\[
\frac{d}{dt}(u) + D_s\left(\frac{\rho^2}{2}\right) = 0,
\]
and
\[
\frac{d}{dt}(t u - x) + D_s\left(\frac{1}{2} t \rho^2 \right) = 0.
\]
Finally, in Euler's  coordinates the conservation
laws (\ref{ek:CL3}) and (\ref{ek:CL4}) become
\[
D_t^e(\rho u) + D_x\left(\frac{\rho^2}{2} + \rho u^2 \right) = 0,
\]
and
\[
D_t^e(\rho (t u - x)) + D_x\left(\frac{1}{2} t \rho^2 + \rho u (t u - x)\right) = 0.
\]

\begin{remark}
Generators~(\ref{ek:syms6}) in Lagrangian mass coordinates read
\[
\def\arraystretch{1.5}
\begin{array}{c}
    Y_1 = \ppartial{t},
    \quad
    Y_2 = \ppartial{s},
    \quad
    Y_3 = \ppartial{x},
    \quad
    Y_4 = t\ppartial{x} + \ppartial{u},
    \\
    Y_5 = 3 t \ppartial{t} + 2 x\ppartial{x} - u \ppartial{u} - 2 \rho \ppartial{\rho},
    \qquad
    Y_6 = 3 s \ppartial{s} + x \ppartial{x} + u \ppartial{u} + 2 \rho \ppartial{\rho}.
\end{array}
\]
\end{remark}

Now we will consider an invariant difference schemes for shallow
water equations in  Lagrangian coordinate system.
Here we change equation (\ref{ek:sw_stetEq}) with constant equals~$1/2$ into the same equation with constant
equals~$1$ as it was used in group classification \cite{bk:KaptsovMeleshko_1D_classf[2018]}
(the same result can be obtained by means of stretching   $x = \alpha \tilde{x}$, where $2\alpha^3 = 1$):
\begin{equation} \label{ek:sw_stetEq2}
p = \rho^2,
\end{equation}
 and then equation (\ref{ek:elSW0}) reads
\begin{equation}\label{ek:elSW}
  x_{tt} - 2 \frac{x_{ss}}{x_s^3} = 0.
\end{equation}

%%%%%%%%%%%%%%%%%%%%%%%%%%%%%%%%%%%%%%%%%%%%%%%%%%%%%%%%%%%%%%%%%%%%%%%%%%%%
%%%%%%%%%%%%%%%%%%%%%%%%%%%%%%%%%%%%%%%%%%%%%%%%%%%%%%%%%%%%%%%%%%%%%%%%%%%%
%%%%%%%%%%%%%%%%%%%%%%%%%%%%%%%%%%%%%%%%%%%%%%%%%%%%%%%%%%%%%%%%%%%%%%%%%%%%

\section{Invariant conservative difference schemes
for the shallow water equations
in Lagrangian coordinates}

\begin{figure}[ht]
\centering
\includegraphics[scale=0.75]{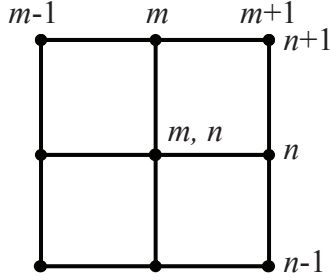}
\caption{9-point stencil}
\label{ek:fig:template_9pt}
\end{figure}

Let us consider invariant difference schemes on the following 9-point stencil
\begin{equation}\label{ek:9ptspace}
(\mathbf{t},\mathbf{s},\mathbf{x}) =
  (
        t, \check{t}, \hat{t};
        s, s_-, s_+;
        x, x_+, x_-,
        \hat{x}, \hat{x}_+, \hat{x}_-,
        \check{x}, \check{x}_+, \check{x}_-
    )
\end{equation}
or, alternatively,
\[
(\mathbf{t},\mathbf{s},\mathbf{x}) =
  (
        t_n, t_{n-1}, t_{n+1};
        s_m, s_{m-1}, s_{m+1};
        x^n_m, x^n_{m+1}, x^{n+1}_{m},
        x^{n+1}_m, x^{n+1}_{m+1}, x^{n+1}_{m-1},
        x^{n-1}_m, x^{n-1}_{m+1}, x^{n-1}_{m-1}
    ),
\]
which is depicted in Fig.~\ref{ek:fig:template_9pt}.

Here and further we will consider the simplest orthogonal regular mesh
\begin{equation}\label{ek:regmesh}
  h^s_- = h^s_+ = h^s = \textrm{const},
\qquad
\tau_- = \tau_+ = \tau = \textrm{const},
\end{equation}
which is invariant under the whole set of the generators~(\ref{ek:syms6})~(see criterion in~\cite{bk:Dorodnitsyn[2011]}).

%\subsubsection{Construction of Invariant difference scheme construction}

\subsection{ The construction of invariant difference scheme}
\label{sec:scheme_construction}

Now we construct an invariant difference scheme for the flat bottom case (\ref{ek:elSW}).

One can easily state that the following difference expression admits
the whole set of the generators~(\ref{ek:syms6}):
\[
    \frac{\tau_-^{4/3}}{(h^s_-)^{1/3}} {\DTM}(x_t).
\]
Hence, we are looking for an invariant scheme having the following divergent form
\begin{equation} \label{ek:scheme_search}
  \frac{\tau_-^{4/3}}{(h^s_-)^{1/3}} \left({\DTM}(x_t) + \DSM(\Phi)\right) = 0,
\end{equation}
where $\underset{\pm\tau}{D}$ and $\underset{\pm s}{D}$ are the difference operators
\[
    \underset{+\tau}{D} = \frac{1 - \STP}{t_{n+1} - t_{n}},
    \quad
    \underset{-\tau}{D} = \frac{\STM - 1}{t_n - t_{n-1}},
    \quad
    \underset{+s}{D} = \frac{1 - \SSP}{s_{m+1} - s_{m}},
    \quad
    \underset{-s}{D} = \frac{\SSM - 1}{s_m - s_{m-1}},
\]
$\underset{\pm\tau}{S}$ and $\underset{\pm s}{S}$ are the difference shift operators that defined as follows
\[
    \underset{\pm\tau}{S}(f(t_n, s_m)) = f(t_n \pm \tau, s_m),
    \qquad
    \underset{\pm s}{S}(f(t_n, s_m)) = f(t_n, s_m \pm h^s),
\]
$x_t=\DTP(x)$ and $x_s=\DSP(x)$ are right difference derivatives with respect to~$t$ and~$s$, $\Phi$ is a function, which approximates $D_s(1/x_s^2)$ to the first order by $h$ and $\tau$.

It is reasonable to consider $\Phi$ as a rational function
\begin{equation}\label{ek:phi_search}
  \Phi(\mathbf{x}) = 1/P_2(\mathbf{x}),
\end{equation}
where $P_2 \neq 0$ --- is a second-order polynomial in points
$\mathbf{x} = \{x, x^+, \hat{x}, \hat{x}^+, \check{x}, \check{x}^+\}$.

Notice that being invariant to the actions of the generators $X_1$--$X_3$
the function~$\Phi(\mathbf{x})$ should depend on the differences:
\[
\def\arraystretch{1.5}
\begin{array}{c}
    t_{n+p+1} - t_{n+p} = \tau_{n+p},
    \quad
    s_{m+q+1} - s_{m+q} = h^s_{m+q},
    \\
    x^{n+p+1}_{m+q} - x^{n+p}_{m+q},
    \quad
    x^{n+p}_{m+q+1} - x^{n+p}_{m+q},
    \quad
    p,q \in \mathbb{Z}.
\end{array}
\]
Therefore, we  consider the following form of the polynomial~$P_2$:
\begin{equation}\label{ek:P2poly}
    P_2(\mathbf{x}) =
        a_{11} x_s^2
        + a_{22} \hat{x}_s^2
        + a_{33} \check{x}_s^2
        + 2 a_{12} x_s\hat{x}_s
        + 2 a_{13} x_s\check{x}_s
        + 2 a_{23} \hat{x}_s\check{x}_s
        + b_1 x_s
        + b_2 \hat{x}_s
        + b_3 \check{x}_s
        + c_1,
\end{equation}
where $a_{ij}$, $b_k$ and $c_1$ are some real constants,
$\hat{x}_s = \STP(x_s)$, and  $\check{x}_s = \STM(x_s)$.

Invariant scheme~(\ref{ek:scheme_search}) must
satisfy the infinitesimal criterions for the invariance, namely
\[
    X_k\left(
        \frac{\tau_-^{4/3}}{(h^s_-)^{1/3}} \left({\DTM}(x_t) + \DSM(\Phi)\right)
    \right)\Bigg|_{(\ref{ek:regmesh})} = 0,
    \qquad
    k = 1,\dots, 6.
\]
After some standard algebraic simplifications one states that
$b_1 = b_2 = b_3 = c_1 = 0$ and the constants~$a_{ij}$ are arbitrary.
Actually, in order to approximate the original equation the following relation must hold
\begin{equation}\label{ek:aij_rel}
  \sum_{i,j} a_{ij} = 1.
\end{equation}

Expression $\DSM(\Phi)$ is divergent one, and it obviously possesses the integrating
factors $1$, $t$ and any function $f(t)$.
As an approximation for the integration factor $x_t$ of
the conservation law of energy~(\ref{CL1}), one can choose the following
difference expression
\begin{equation}\label{ek:energy_mult}
  \Lambda = \frac{x_t + \check{x}_t}{2}.
\end{equation}
Following the difference analog of the direct method we demand
\begin{equation}\label{VDdirect}
  \mathcal{E}\left( \Lambda ({\DTM}(x_t) + \DSM(\Phi)) \right)\Bigg|_{(\ref{ek:regmesh})} = 0,
\end{equation}
where
\[
\mathcal{E} =
    \sum_{k=-\infty}^{+\infty}
    \sum_{l=-\infty}^{+\infty}
    {\STP}^k{\SSP}^l
    \ppartial{x^{n-k}_{m-l}},
\]
is the variational Euler operator (see \cite{bk:Dorodnitsyn[2011]}).

Substituting~(\ref{ek:P2poly}) and~(\ref{ek:energy_mult}) into~(\ref{VDdirect})
and carrying out the summation of the resulting rational expressions,
one gets the fraction
\[
    \mathcal{E}\left( \Lambda ({\DTM}(x_t) + \DSM(\Phi)) \right)\Bigg|_{(\ref{ek:regmesh})}
    = \frac{P_{54}}{P_{\cdots}} = 0,
\]
where $P_{54}$ is a polynomial of degree~54 in terms of~20 points
\[
    x^{n+i}_{m+j}, \quad i=-2,\dots,2, \quad  j=-1,\dots,2,
\]
and by $P_{\cdots}$ we denote some nonzero polynomial defined on the same set of points.
Then, considering the leading terms of the polynomial~$P_{54}$, one sees that
the following equation must be satisfied by the coefficients
\begin{equation}\label{ek:aswitch1}
  a_{11}a_{22}a_{33}(a_{33} - a_{22}) = 0.
\end{equation}
Let us choose the simplest solution~$a_{11} = a_{22} = a_{33} = 0$.
Substituting the solution into the polynomial~$P_{54}$ and considering the leading terms again, one gets
\begin{equation}\label{ek:aswitch2}
  a_{13}a_{23}(a_{13} - a_{23}) = 0.
\end{equation}
Putting~$a_{13}=a_{23}=0$, one derives the following form of the function sought
\[
    \Phi = \frac{1}{P_2} = \left(a_{23} \hat{x}_s \check{x}_s\right)^{-1},
\]
which satisfies~(\ref{VDdirect}) for any value of the coefficient~$a_{23}$.
Taking~(\ref{ek:aij_rel}) into account, one puts~$a_{23} = 1$ and
finally arrives at the following finite-difference scheme on regular orthogonal mesh:
\large
\begin{equation} \label{ek:scheme_lagr}
  \displaystyle
  \def\arraystretch{1.75}
  \begin{array}{c}
        F = x_{t\check{t}}
      + \frac{1}{h^s_-} \left(
        (\hat{x}_s \check{x}_s)^{-1}
        - (\hat{x}_{\bar{s}} \check{x}_{\bar{s}})^{-1}
      \right)
      = 0,
      \\
      \tau_+ = \tau_-, \qquad
      h^s_+ = h^s_-,
  \end{array}
\end{equation}
\normalsize
where $x_{\bar{s}} = \underset{-h}{D}(x)$ and $x_{t\check{t}} = \underset{-\tau}{D}(x_t)$.
The scheme (\ref{ek:scheme_lagr}) approximates equation (\ref{ek:elSW}) up to $O((h^s)^2 + \tau^2)$.

\begin{remark}
One can check that choosing another solutions of equation~(\ref{ek:aswitch1})~(or equation~(\ref{ek:aswitch2})) finally leads to
either the solution~(\ref{ek:scheme_lagr}),
or the trivial solution~$P_2 = 0$~(which is prohibited by~(\ref{ek:phi_search})).
\end{remark}

\begin{remark}
One can rewrite (\ref{ek:scheme_lagr}) in the following ``viscosity'' form:
\begin{equation}\label{ek:viscous_scheme}
  \displaystyle
  x_{t\check{t}}
  + \frac{1}{h^s_-} \left(
    (\hat{x}_s \check{x}_s)^{-1}
    - (\hat{x}_{\bar{s}} \check{x}_{\bar{s}})^{-1}
  \right)
  + \mu \frac{\hat{x}_{s\bar{s}} + \check{x}_{s\bar{s}}}{2} = 0,
\end{equation}
where $|\mu | \ll (h^s)^2$ is the viscosity factor.
%One can show all the obtained below conservation laws of the scheme are still hold in this case.
\end{remark}

\begin{remark}
Apparently, (\ref{ek:scheme_lagr}) is not the only scheme
that could be obtained by the procedure similar to performed above.
%However, it seems to be the simplest one, and that is enough for further purposes.
For example, one may generalize the scheme (\ref{ek:viscous_scheme}), assuming
\[
  x_t = \DTP(w(x)), \quad x_s = \DSP(w(x)),
\]
where $w$ is a function. Substituting into (\ref{ek:viscous_scheme})
and performing series expansion, one gets
\[
    w^\prime (x_{tt} - 2 x_s^{-3} x_{ss}) + w^{\prime\prime} (x_t^2 - 2 x_s^2) + O(\tau^2 + h^2).
\]
Then, the scheme can be represented in the following generalized form
\[
  \displaystyle
  w(x)_{t\check{t}}
  + \frac{1}{h^s_-} \left(
    (w(\hat{x})_s w(\check{x})_s)^{-1}
    - (w(\hat{x})_{\bar{s}} w(\check{x})_{\bar{s}})^{-1}
  \right)
  + \mu \frac{w(\hat{x})_{s\bar{s}} + w(\check{x})_{s\bar{s}}}{2} = 0,
\]
where
\[
    w(z) = \nu \varphi(z) + z + c,
\]
and $c$ is constant, $\varphi$ is an arbitrary function,
and $\nu$ is a coefficient of order~$O((h^s)^2 + \tau^2)$.
\end{remark}

\subsection{Invariant representation of the scheme (\ref{ek:scheme_lagr})}
\label{ek:sec:delta_case_flat}

Consider invariants of the symmetries (\ref{ek:syms6})
in the space (\ref{ek:9ptspace}).
There are $15 - 6 = 9$ difference invariants:
\begin{equation}\label{ek:shallow_water_invariants}
    \def\arraystretch{1.75}
    \begin{array}{c}
        I_1 = \frac {h^s_+}{h^s_-},
        \qquad
        I_2 = \frac {\tau_+}{\tau_-},
        \\
        I_3 = {\frac {{ x}-{ x_-}}{\omega}},
        \quad
        I_4 = {\frac {{ x_+}-{ x}}{\omega}},
        \quad
        I_5 = {\frac {{ \check{x}_+}-{ \check{x}}}{\omega}},
        \quad
        I_6 = {\frac {{ \check{x}}-{ \check{x}_-}}{\omega}},
        \\
        I_7 = {\frac {{ \tau_+}\, \left( { x}-{ \check{x}} \right)
        + { \tau_-} \left( {x}-{ \hat{x}} \right) }{\omega\tau_-}},
        \qquad
        I_8 = {\frac {{ \tau_+}\, \left( { x}-{ \check{x}} \right)
            + { \tau_-} \left( { x}-{ \hat{x}_+} \right) }{\omega\tau_-}},
        \\
        I_9 = {\frac { { \tau_+}\,\left( { x}-{ \check{x}} \right) + { \tau_-} \left( { x}-{ \hat{x}_-} \right)
        }{\omega\tau_-}},
\end{array}
\end{equation}
where
\[
    \omega^3 = \tau_-^2 h^s_-, \qquad
    h^s_+ = s_+ - s, \qquad
    h^s_- = s - s_-.
\]

The scheme (\ref{ek:scheme_lagr}) is invariant to the actions of the
whole 6-parametric group (\ref{ek:syms6}):
\[
  X_k F |_{(\ref{ek:scheme_lagr})} = 0, \qquad k = 1, ..., 6.
\]
Thus, the scheme (\ref{ek:scheme_lagr}) can be represented in terms of the invariants (\ref{ek:shallow_water_invariants}),
%\[
%  (\tau_-)^{4/3} (h^s_-)^{-1/3},
%\]
namely,
\large
\[
\def\arraystretch{1.75}
\begin{array}{c}
  \frac{(\tau_-)^{4/3}}{(h^s_-)^{1/3}}\left(
        x_{t\check{t}}
            + \DSM\left((\hat{x}_s \check{x}_s)^{-1}\right)
  \right)
  =
  \frac {
    I_2 - I_6 I_7\, (I_7 - I_9)
  }
  {
    I_2 I_6\, (I_7 - I_9)
  }
  + \frac{
    (I_1)^{2}
  }
  {
    I_5\, (I_7 - I_8)
  } = 0,
  \\
  I_1 = 1,
  \qquad
  I_2 = 1.
\end{array}
\]
\normalsize

\begin{remark}
Invariant form of the ``viscosity'' scheme (\ref{ek:viscous_scheme}) is the following
\[
\def\arraystretch{1.75}
\begin{array}{c}
  \frac {
    I_2 - I_6 I_7\, (I_7 - I_9)
  }
  {
    I_2 I_6\, (I_7 - I_9)
  }
  + \frac{
    (I_1)^{2}
  }
  {
    I_5\, (I_7 - I_9)
  }
  + \frac{\mu \alpha^2}{2}
  \left(
    \frac{I_5 + I_7 - I_9}{I_1}
    + I_7 - I_9 - I_6
  \right)
  = 0,
  \\
  I_1 = 1,
  \qquad
  I_2 = 1,
\end{array}
\]
where $\alpha = \tau_- / {h^s_-}$ (on regular mesh,
one can consider $\alpha$ as constant).
\end{remark}

%\begin{remark}
%Схема (\ref{ek:viscous_scheme}) может быть дополнительно обобщена с сохранением основных ЗС.
%Будем использовать следующие формы разностных производных:
%\begin{equation}
%  x_t = \DTP(w(x)), \quad x_s = \DSP(w(x)),
%\end{equation}
%где $w$ -- некоторая функция. Подставляя в схему (\ref{ek:viscous_scheme}) и разлагая в ряд, получим:
%\[
%    w^\prime (x_{tt} - 2 x_s^{-3} x_{ss}) + w^{\prime\prime} (x_t^2 - 2 x_s^2) + O(\tau^2 + h^2).
%\]
%Отсюда можно сделать вывод, что схема может быть представлена в обобщённом виде:
%  \begin{equation}
%  \displaystyle
%  w(x)_{t\check{t}}
%  + \frac{1}{h^s_-} \left(
%    (w(\hat{x})_s w(\check{x})_s)^{-1}
%    - (w(\hat{x})_{\bar{s}} w(\check{x})_{\bar{s}})^{-1}
%  \right)
%  + \mu \frac{w(\hat{x})_{s\bar{s}} + w(\check{x})_{s\bar{s}}}{2} = 0,
%\end{equation}
%где
%\[
%    w(z) = \nu \varphi(z) + z + c,
%\]
%$c$ -- некоторая константа, $\varphi$ -- произвольная функция, $\nu$ --- коэффициент порядка $O(h^2 + \tau^2)$.
%При этом интегрирующий множитель $(x_t + \check{x}_t)/2$ перейдет в множитель $(w(x)_t + w(\check{x})_t)/2$.
%\end{remark}

\subsection{Conservation laws of the scheme (\ref{ek:scheme_lagr})}

It was stated in Section~\ref{sec:scheme_construction} that the scheme
(\ref{ek:scheme_lagr}) has the following set of
integration multipliers:
\begin{equation}\label{ek:scheme_mults}
    \Lambda_1 = 1,
    \quad
    \Lambda_2 = t,
    \quad
    \Lambda_3 = \frac{x_t + \check{x}_t}{2},
\end{equation}
where the last multiplier was given by the formula (\ref{ek:energy_mult}).
The multipliers (\ref{ek:scheme_mults}) allow one to writhe
the following conservation laws of the scheme (\ref{ek:scheme_lagr})
in Lagrangian coordinates.
\begin{enumerate}
    \item The continuity equation
    \begin{equation} \label{ek:mixed_derivs_cl}
        {\DTM}(\hat{x}_s) - \DSM(x_t^+) = 0,
    \end{equation}
    which is just the difference analogue
    of the condition (\ref{x_commute}).

  \item
  Conservation of momentum:
  \begin{equation}
    \Lambda_1 = 1, \qquad
    {\DTM}(x_t) + \DSM\left(
        (\hat{x}_s \check{x}_s)^{-1}
    \right) = 0,
  \end{equation}
  which corresponds to (\ref{ek:CL3}).

  \item
  Center of mass conservation:
  \begin{equation}
    \Lambda_2 = t, \qquad
    {\DTM}(t x_t - x) + \DSM\left(
        t (\hat{x}_s \check{x}_s)^{-1}
    \right) = 0,
  \end{equation}
  which corresponds to (\ref{ek:CL4}).

  \item
  Conservation of energy:
  \begin{equation}
    \Lambda_3 = \frac{x_t + \check{x}_t}{2}, \qquad
    \frac{1}{2} {\DTM}(
        x_t^2 + x_s^{-1} + \hat{x}_s^{-1}
    )
    + \frac{1}{2} \DSM\left(
        (x_t^+ + \check{x}_t^+)
        (\hat{x}_s \check{x}_s)^{-1}
    \right) = 0,
  \end{equation}
  which corresponds to (\ref{CL1}).
\end{enumerate}

\begin{remark}
  Conservation laws for the scheme (\ref{ek:viscous_scheme}) can be written as follows.
  \begin{enumerate}
  \item Conservation of momentum:
  \begin{equation}
    {\DTM}(x_t) + \DSM\left(
        (\hat{x}_s \check{x}_s)^{-1}
        + \mu\frac{\hat{x}_s + \check{x}_s}{2}
    \right) = 0.
  \end{equation}

  \item Center of mass conservation:
  \begin{equation}
    {\DTM}(t x_t - x) + \DSM\left(
        t (\hat{x}_s \check{x}_s)^{-1}
        + t \mu \frac{\hat{x}_s + \check{x}_s}{2}
    \right) = 0.
  \end{equation}

  \item Conservation of energy:
  \[
    \def\arraystretch{1.5}
    \begin{array}{c}
    \frac{1}{2} {\DTM}(
        x_t^2 + x_s^{-1} + \hat{x}_s^{-1}
        + \mu \tau \left(
            x x_+ + \hat{x}\hat{x}_+
            - 2 (x^2 + \hat{x}^2)
        \right)
    ) {}\\
    {} + \frac{1}{2} \DSM\left(
        (x_t^+ + \check{x}_t^+)
        (\hat{x}_s \check{x}_s)^{-1}
        + \mu h^s (\hat{x}\check{x}_+ - \check{x}\hat{x}_+)
    \right) = 0.
    \end{array}
  \]
\end{enumerate}
\end{remark}

\subsection{Schemes and their conservation laws in Lagrangian mass coordinates}

In the present section, through difference transformations of type~(\ref{pot}),
we construct two invariant conservative difference schemes in Lagrangian mass coordinates,
which correspond to the scheme~(\ref{ek:scheme_lagr}).
Using naive difference transformations of type~(\ref{pot}), one can derive a scheme on three time layers,
which is the first scheme presented here. The second scheme is defined on two time layers,
and some more sophisticated transformations are required in that case.

\subsubsection{Three-level scheme}

\begin{figure}[h]
\centering
\includegraphics[scale=0.75]{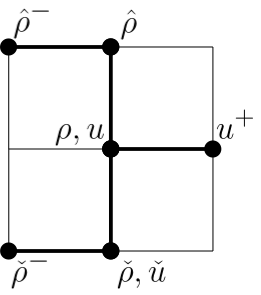}
\caption{Three-level scheme difference template}
\label{ek:fig:template_t3}
\end{figure}

Here we use the following difference analog of the transformation~(\ref{pot})
\begin{equation} \label{ek:contact_transform_lagr}
  x_t = u, \qquad x_s = \rho^{-1}.
\end{equation}
The scheme~(\ref{ek:scheme_lagr}) becomes
\begin{equation} \label{ek:scheme_mass_coord}
    \def\arraystretch{1.75}
    \begin{array}{c}
    \left( \frac{1}{\rho} \right)_t = u_s,
    \\
      {\DTM}(u) + \DSM\left(
        \hat{\rho} \check{\rho}
    \right) = 0, \\
    h^s_+ = h^s_-, \quad
    \tau_+ = \tau_-.
    \end{array}
\end{equation}
The corresponding difference template is shown in the Figure~\ref{ek:fig:template_t3}

%\begin{remark}
We remind that in the case of the shallow water equations
the physical meaning of variable $\rho$
is thickness of a water column above the bottom.
%\end{remark}

Using the following approximation of ``state'' equation (\ref{ek:sw_stetEq2})
\begin{equation} \label{ek:state_eq_raw}
  p = \rho^\gamma = \rho^2,
  %\qquad
  %\varepsilon = \frac{p}{(\gamma - 1) \rho} = p / \rho.
\end{equation}
one can represent the scheme (\ref{ek:scheme_mass_coord}) in the form
\[
    \def\arraystretch{1.75}
    \begin{array}{c}
    \left( \frac{1}{\rho} \right)_t = u_s,
    \\
      {\DTM}(u) + \DSM\left(
        \sqrt{\hat{p} \check{p}}
    \right) = 0, \\
    h^s_+ = h^s_-, \quad
    \tau_+ = \tau_-.
    \end{array}
\]
The conservation laws of the latter scheme are the following.
% THE LIST OF LAWS' NAMES:
%Conservation of mass:
%The conservation law of momentum:
%Center of mass conservation:
%The conservation law of energy:
\begin{enumerate}
    \item Conservation of mass:
    \[
    \left( \frac{1}{\rho} \right)_t = u_s,
    \]

  \item The conservation law of momentum:
  \[
    \Lambda = 1, \qquad
    {\DTM}(u) + \DSM\left(
        \hat{\rho} \check{\rho}
    \right)
    = {\DTM}(u) + \DSM\left(
        \sqrt{\hat{p} \check{p}}
    \right)
    = 0.
  \]

  \item Center of mass conservation:
  \[
    \Lambda = t, \qquad
    {\DTM}(t u - x) + \DSM\left(
        t \hat{\rho} \check{\rho}
    \right)
    = {\DTM}(t u - x) + \DSM\left(
        t \sqrt{\hat{p} \check{p}}
    \right)
    = 0.
  \]

  \item The conservation law of energy:
  \[
    \Lambda = \frac{u + \check{u}}{2}, \qquad
    \frac{1}{2} {\DTM}\left(
        u^2 + \frac{p}{\rho} + \frac{\hat{p}}{\hat{\rho}}
    \right)
    + \frac{1}{2} \DSM\left(
        (u^+ + \check{u}^+)
        \sqrt{\hat{p} \check{p}}
    \right) = 0.
  \]

\end{enumerate}

\subsubsection{Two-level scheme}
\label{ek:sec:2layer_scheme}
It is of our interest to construct two-level difference schemes, as far one can perform
numerical calculations much easier with the help of such a scheme.
In the space of finite--differences one has a freedom to choose approximations of the  relations~(\ref{pot}) and of the ``state equation''~(\ref{ek:sw_stetEq2}).

\begin{figure}[h]
\centering
\includegraphics[scale=0.75]{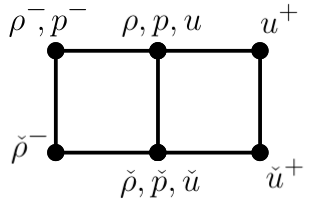}
\caption{Two-level scheme difference template}
\label{ek:fig:template_t2}
\end{figure}

Here, instead of~(\ref{ek:contact_transform_lagr}), we use the following transformation
\begin{equation} \label{ek:contact_transform_lagr2}
  \check{x}_s + x_s = \frac{2}{\check{\rho}}, \qquad
  %{x}_s + \check{x}_s = 2 \check{\rho}^{-1}, \qquad
  x_t = u,
\end{equation}
and the following implicit approximation of ``state'' equation~(\ref{ek:sw_stetEq2})
\begin{equation} \label{ek:state_eq_raw2}
    \frac{1}{\sqrt{\check{p}}} + \frac{1}{\sqrt{p}} = \frac{2}{\check{\rho}}
    \quad \Leftrightarrow \quad
    x_s = \frac{1}{\sqrt{p}}.
\end{equation}
Then, the scheme (\ref{ek:scheme_mass_coord}) becomes
\begin{equation} \label{ek:scheme_mass_coord2}
    \def\arraystretch{1.75}
    \begin{array}{c}
    {\DTM}\left( \frac{1}{\rho} \right) - \DSM\left(
        \frac{u^+ + \check{u}^+}{2}
    \right) = 0,
    \\
      {\DTM}(u) + \DSM\left(
        Q
    \right) = 0, \\
    h^s_+ = h^s_-, \quad
    \tau_+ = \tau_-,
    \end{array}
\end{equation}
where
\begin{equation} \label{ek:flux_Q}
  \frac{1}{Q} = \frac{4}{\rho \check{\rho}}
    - \frac{2}{\sqrt{p}}\left( \frac{1}{\rho} + \frac{1}{\check{\rho}} \right)
     + \frac{1}{p}.
\end{equation}
Here the scheme and the ``state'' equation are written
in the points of \emph{two} difference time layers.
The corresponding difference template is shown on the Figure~\ref{ek:fig:template_t2}.
The scheme (\ref{ek:scheme_mass_coord2}) approximates equation (\ref{ek:elSW}) up to $O(h+\tau)$.

The conservation laws of the scheme (\ref{ek:scheme_mass_coord2}) are the following.
\begin{enumerate}
    \item Conservation of mass\footnote{To get this conservation law one should instead of (\ref{ek:mixed_derivs_cl}) get the sum of (\ref{ek:mixed_derivs_cl}) and
        the shifted one, i.e.,
        \[{\DTM}(\hat{x}_s + x_s) - \DSM(x_t^+ + \check{x}_t^+) = 0.\]}:
    \[
    {\DTM}\left( \frac{1}{\rho} \right) - \DSM\left(
        \frac{u^+ + \check{u}^+}{2}
    \right) = 0.
    \]

  \item The conservation law of momentum:
  \[
    \Lambda = 1, \qquad
%    {\DTM}(u) + \DSM\left(
%        \hat{\rho} \check{\rho}
%    \right) =
    {\DTM}\left( u \right) + \DSM\left(Q\right) = 0.
  \]

  \item Center of mass conservation:
  \[
    \Lambda = t, \qquad
    %{\DTM}(t u - x) + \DSM\left(
%        t \hat{\rho} \check{\rho}
%    \right) =
    {\DTM}(t u - x) + \DSM\left(t Q \right)
    = 0.
  \]

  \item The conservation law of energy:
  \[
    \displaystyle
    \Lambda = \frac{u + \check{u}}{2}, \qquad
        {\DTM}\left(
            \frac{u^2}{2} + \frac{p}{2 \sqrt{p} - \rho}
    \right)
    + \DSM\left(
        \frac{u^+ + \check{u}^+}{2} \, Q
    \right) = 0.
  \]

\end{enumerate}

\begin{remark}
Here ``viscosity'' term of the scheme (\ref{ek:viscous_scheme}) becomes:
\[
  \displaystyle
  \mu \frac{\hat{x}_{s\bar{s}} + \check{x}_{s\bar{s}}}{2}
  = \frac{\mu}{h^s}\left(
    \frac{1}{\rho} + \frac{1}{\check{\rho}} - \frac{1}{\sqrt{p}}
    -\frac{1}{\rho^-} - \frac{1}{\check{\rho}^-} + \frac{1}{\sqrt{p^-}}
  \right).
\]
Apparently, the physical meaning of this term is lost in Lagrangian mass coordinates.
\end{remark}

\subsection{Some invariant solutions of the scheme}

In the present section we consider some invariant solutions of the scheme~(\ref{ek:scheme_lagr}).
It seems that in Lagrangian coordinates the scheme has a wider class of invariant solutions
than in Eulerian representation~(in particular there is no travelling wave solutions in Eulerian coordinates).

\subsubsection{Example 1. Travelling wave solution}

Consider travelling wave solutions
\[
    x = \psi(\lambda),
    \qquad
    \lambda = s - \alpha t,
\]
which correspond to the subalgebra~$X_{1,5} = X_1 + \alpha X_5 = \partial_t + \alpha \partial_s$.
Then, equation~(\ref{ek:elSW0}) becomes
\[
    \psi^{\prime\prime}(\alpha^2 (\psi^\prime)^3 - 2) = 0.
\]
Further we consider the simplest form of travelling wave solution that is
\begin{equation} \label{ex1_invsol}
  x = \lambda = s - \alpha t.
\end{equation}

Constructing the scheme on the subgroup corresponding to the generator~$X_{1,5}$,
one should mention that the difference mesh spacing~$\Delta\lambda$ along the~$\lambda$-axis
should be matched with the original mesh spacings~$h$ and~$\tau$~\cite{bk:Dorodnitsyn[2011]}.
This is achieved under conditions
\[
    \alpha = \frac{h}{\tau},
    \qquad
    \Delta\lambda = h.
\]
Then, scheme~(\ref{ek:scheme_lagr}) reduces to the following
\[
\displaystyle
  \def\arraystretch{1.75}
  \scalebox{1.25}{$
  \begin{array}{c}
        \alpha^2 \psi_{\lambda\bar{\lambda}}
         + \left[
                \underset{+\lambda}{S}\left(\frac{1}{\psi_{\lambda}}\right)
                \underset{-\lambda}{S}\left(\frac{1}{\psi_{\lambda}}\right)
         \right]_{\bar{\lambda}} = 0,
        \\
        \lambda_+ - \lambda = \lambda - \lambda_- = \Delta\lambda,
    \end{array}
    $}
\]
where $\psi = \psi(\lambda)$.
The latter scheme possesses invariant solution~(\ref{ex1_invsol}),
\[
    x = \psi(\lambda) = \lambda,
\]
which becomes trivial in Eulerian coordinates:
\[
  u = -\alpha,
  \qquad
  \rho = 1.
\]

\subsubsection{Example 2. Nonuniform dilation in the $(x,s)$-space}

Consider subalgebra~$X_6 = 3 s \partial_s + x \partial_x$
and the corresponding invariant solution of equation~(\ref{ek:elSW0}):
\begin{equation} \label{exactsol}
    x = s^{1/3} \psi,
    \qquad
    u = s^{1/2} \psi^\prime,
    \qquad
    \rho = 3 s^{2/3} / \psi,
\end{equation}
where $\psi = \psi(t)$.

Equation~(\ref{ek:elSW0}) reduces to the form
\begin{equation} \label{reduced}
    \psi^2 \psi^{\prime\prime} + 12 = 0.
\end{equation}
One can find the following particular exact solution of the latter equation:
\begin{equation} \label{partsol}
  x = s^{1/3} \psi(t) = (54 s t^2)^{1/3}.
\end{equation}

The generator~$X_6$ holds orthogonality of the mesh.
Therefore, further we consider the scheme on
an orthogonal mesh without requiring it to be a uniform one.

Then, the first equation of the scheme~(\ref{ek:elSW0}) possesses the form
\begin{equation} \label{reduced_scheme}
    \displaystyle
    \scalebox{1.25}{$
        \hat{\psi}\check{\psi} \psi_{t\hat{t}}
        + \frac{1}{s^{1/3}(s - s_-)} \, \left(
            \left(\frac{s_+ - s}{s_+^{1/3} - s^{1/3}}\right)^2
            - \left(\frac{s_- - s}{s_-^{1/3} - s^{1/3}}\right)^2
        \right) = 0,
    $}
\end{equation}
and we do not set the mesh equations for now.

According to~(\ref{exactsol}), one can split scheme~(\ref{reduced_scheme}) into two separate equations:
\begin{equation}\label{split1}
  \hat{\psi}\check{\psi} \psi_{t\hat{t}} = K = \text{const},
\end{equation}
and
\begin{equation}\label{split2}
  \frac{1}{s^{1/3}(s - s_-)} \, \left(
            \left(\frac{s_+ - s}{s_+^{1/3} - s^{1/3}}\right)^2
            - \left(\frac{s_- - s}{s_-^{1/3} - s^{1/3}}\right)^2
        \right) = -K = \text{const}.
\end{equation}
By virtue of~(\ref{reduced}), one can set $K = \delta - 12$, where $0 < |\delta| \ll 1$ is a constant
related to the mesh density.

\medskip

First, consider equation~(\ref{split1}) on a uniform lattice.
One can obtain its first integral
\begin{equation}
  \psi_t^2 + K \left(
    \frac{1}{\hat{\psi}} + \frac{1}{\psi}
  \right) = C_1 = \text{const},
\end{equation}
which corresponds to the integrating factor
~$(\hat{\psi} - \check{\psi})/({\hat{\psi}\check{\psi}})$.
Then, one can writh the following implicit solution of the Cauchy problem
\begin{equation}
  (\psi(t_n) - \psi_0)^2 + t_n^2 K  \left(
    \frac{1}{\psi(t_n)} + \frac{1}{\psi_0}
  \right) - t_n^2 C_1 = 0,
\end{equation}
\[
    t_n = n \tau,
    \qquad
    n \in \mathbb{Z},
\]
where $\psi_0 = \psi(0)$ is the value of the function at the initial time.

\begin{remark}
Notice that in some specific cases equations of type~(\ref{split1})
can be linearized~\cite{bk:DorodLinearization}.
\end{remark}

Next, consider equation~(\ref{split2}).
One can find a particular (invariant) solution of~(\ref{split2}).
Consider an invariant grid
\begin{equation}
\frac{s_+}{s} = \frac{s}{s_-},
\end{equation}
which possesses the first integral
\begin{equation}\label{eqh0}
 s_+ = \kappa^3 s, \qquad \kappa = \text{const},
\end{equation}
One can integrate the latter equation and get
\begin{equation}\label{eqh}
  s_m = \kappa^{3 m} s_0, \qquad m \in \mathbb{Z},
\end{equation}
where $s_0$ is an initial constant value.

It is easy to verify that equation~(\ref{eqh})
becomes a particular solution of~(\ref{split2}) under the following condition
\begin{equation}\label{eqk}
(\kappa^2 + \kappa + 1)(\kappa^2 + 1)(\kappa + 1) = (12 - \delta) \kappa^4.
\end{equation}
The latter equation has (assuming~$\delta$ is small enough) three real positive roots.
Evidently, for~$\delta \rightarrow 0$ one or more of the solutions~$\kappa$ have values close to~$1$.

\medskip

In Lagrangian mass coordinates, functions~$\rho$ and $u$ are added to the solution.
In the simplest case~($u = x_t$, $\rho = 1/x_s$) they are
\begin{equation}
    x = s^{1/3} \psi,
    \qquad
  u = s^{1/3} \psi_t,
  \qquad
  \rho = \frac{s_+ - s}{s_+^{1/3} - s^{1/3}} \, \frac{1}{\psi}.
\end{equation}
On lattice equation~(\ref{eqh0}) the latter solution becomes
\begin{equation}\label{extru}
    x = s^{1/3} \psi,
    \qquad
  u = s^{1/3} \psi_t,
  \qquad
  \rho = (\kappa^2 + \kappa + 1) \frac{s^{2/3}}{\psi}.
\end{equation}
It corresponds to the solution of the original differential equation
for~$\kappa \rightarrow 1$.

\subsubsection*{Particular exact solution}

Now, consider the scheme on a nonuniform mesh and rewrite
equation~(\ref{split1}) in the following expanded form:
\begin{equation}\tag{\ref{split1}}
  \frac{\psi(\hat{t})\psi(\check{t})}{t - \check{t}}\left[
    \frac{\psi(\hat{t}) - \psi(t)}{\hat{t} - t}
    - \frac{\psi(t) - \psi(\check{t})}{t - \check{t}}
  \right] = \delta - 12.
\end{equation}
Substituting invariant solution
\begin{equation}
  \hat{t} = \mu^3 t.
\end{equation}
into~(\ref{split1}), one gets
\begin{equation} \label{teqn}
  \frac{\mu^3 \psi(\mu^3 t)\psi(t/\mu^3)}{(\mu^3 - 1)^2 \, t^2}\left[
    \psi(\mu^3 t) - (1 + \mu^3) \psi(t) + \psi(t/\mu^3)
  \right] = \delta - 12.
\end{equation}
Let us seek for an invariant solution of the form
\[
  \psi(z) = B z^q, \qquad B, q = \text{const}.
\]
Analyzing the left side of equation~(\ref{teqn}), one concludes that $q =2/3$, i.~e.,
\begin{equation}
  \psi(z) = B z^{2/3}, \qquad B = \text{const},
\end{equation}
which corresponds to the solution~(\ref{partsol}).
Therefore, in order to obtain an exact solution, one set
\[
    B = 54^{1/3}.
\]
Substituting into~(\ref{teqn}), one gets the following mesh constraint
\[
    54 \mu^3 (\mu + 1) = (12 - \delta)(\mu^2 + \mu + 1)^2.
    %54 \mu (\mu^4 + \mu + 1) = (12 - \delta)(\mu - 1)(\mu^2 + \mu + 1)^2.
\]
Comparing with~(\ref{eqk}), one obtains a relation connecting the parameters that
specify the grid densities with respect to~$t$ and~$s$:
\begin{equation}
  \frac{(\kappa^2 + \kappa + 1)(\kappa^2 + 1)(\kappa + 1)}{\kappa^4}
  = \frac{54 \mu^3 (\mu + 1)}{(\mu^2 + \mu + 1)^2}.
  %\frac{54 \mu (\mu^4 + \mu + 1)}{(\mu - 1)(\mu^2 + \mu + 1)^2}.
\end{equation}
Hence, one gets the following particular \textit{exact} solution:
\begin{equation}
    \def\arraystretch{1.75}
    \begin{array}{l}
    x^n_m = (54 s_m t_n^2)^{1/3},
    \\
    t_n = \mu^{3 n} t_0,
    \qquad
    s_m = \kappa^{3 m} s_0,
    \qquad
    m,n \in \mathbb{Z},
    \\
    \frac{(\kappa^2 + \kappa + 1)(\kappa^2 + 1)(\kappa + 1)}{\kappa^4}
    = \frac{54 \mu^3 (\mu + 1)}{(\mu^2 + \mu + 1)^2},
    %\frac{54 \mu (\mu^4 + \mu + 1)}{(\mu - 1)(\mu^2 + \mu + 1)^2}.
    \end{array}
\end{equation}
where $t_0$ and $s_0$ are the initial data of the Cauchy problem.

Solution~(\ref{extru}) for the functions $\rho$ and $u$ is
\begin{equation}
  u^n_m = \frac{1-\mu^2}{1-\mu^3} \left(\frac{54 \, s_m}{t_n}\right)^{1/3},
  \qquad
  \rho^n_m = \frac{\kappa^2 + \kappa + 1}{54^{1/3}} \left(\frac{s_m}{t_n}\right)^{2/3}.
\end{equation}

\subsection{Schemes for an arbitrary bottom case}

In this case equation (\ref{ek:elSW}) has the form
\begin{equation}\label{ek:elSW_bottom}
  x_{tt} - 2 \frac{x_{ss}}{x_s^3} - H^\prime(x) = 0,
\end{equation}
where $H(x)$ is a bottom profile function.

\subsubsection{Case $H^\prime(x) \neq 0$.}

In this general case, equation~(\ref{ek:elSW_bottom}) is not a divergent one
and its invariance strongly depends on the specific type of the function~$H(x)$.
In case~$H(x)$ is arbitrary, equation~(\ref{ek:elSW_bottom}) only admits
two generators,
\[
    X_1 = \frac{\partial}{\partial t}
    \quad
    \text{and}
    \quad
    X_2 = \frac{\partial}{\partial s}.
\]
In finite-difference case there is no obvious way to approximate the derivative~$H^\prime(x)$.
One can consider separately two different representations of this derivative, namely
\begin{equation}\label{derivHt}
    H^\prime(x) = \frac{1}{x_t} (H^\prime(x) \,x_t) = \frac{1}{x_t} \, D_t(H(x))
\end{equation}
and
\begin{equation}\label{derivHts}
    H^\prime(x) = \frac{1}{x_s} (H^\prime(x) \,x_s) = \frac{1}{x_s} \, D_s(H(x)).
\end{equation}

According to the representation~(\ref{derivHt}),
scheme~(\ref{ek:scheme_lagr}) can be generalized as follows
\begin{equation} \label{ek:scheme_lagr_bottom}
  \displaystyle
  \def\arraystretch{1.75}
  \scalebox{1.25}{$
  \begin{array}{c}
  x_{t\check{t}} + \DSM\left(
    (\hat{x}_s \check{x}_s)^{-1}
  \right)
    -\frac{1}{x_t + \check{x}_t}\left( {\DTM}{H(x)} + \DTP{H(x)} \right)
  = 0,
  \\
  \tau_+ = \tau_-,
  \qquad
  h^s_+ = h^s_-,
  \end{array}
  $}
\end{equation}
where
\[
\frac{{\DTM}{H(x)} + \DTP{H(x)}}{x_t + \check{x}_t}
= \frac{2}{x_t + \check{x}_t} \left[ \frac{{\DTM}{H(x)} + \DTP{H(x)}}{2} \right]
\sim \frac{1}{x_t^\prime} [ H_x^\prime x_t^\prime ]
\sim H^\prime(x).
\]

Scheme (\ref{ek:scheme_lagr_bottom}) only has
the conservation law of mass~(\ref{ek:mixed_derivs_cl})
and the conservation law of energy:
\[
  %\Lambda = \frac{x_t + \check{x}_t}{2}, \qquad
    {\DTM}(
        x_t^2 + x_s^{-1} + \hat{x}_s^{-1}
        - H(x) - H(\hat{x})
    )
    + \DSM\left(
        (x_t^+ + \check{x}_t^+)
        (\hat{x}_s \check{x}_s)^{-1}
    \right) = 0.
\]

Using (\ref{ek:contact_transform_lagr2}), (\ref{ek:state_eq_raw2}), one obtains
the following representation of the latter conservation law
%of the scheme (\ref{ek:scheme_lagr_bottom})
in Lagrangian mass coordinates
\[
    \displaystyle
    %\Lambda = \frac{u + \check{u}}{2}, \qquad
        \DTM\left(
            \frac{u^2}{2} + \frac{p}{2 \sqrt{p} - \rho}
            - \frac{H(x) + H(x + \tau u)}{2}
    \right)
    + \DSM\left(
        \frac{u^+ + \check{u}^+}{2} \, Q
    \right) = 0,
\]
where $Q$ is given by the formula (\ref{ek:flux_Q}).

\bigskip

Using scheme~(\ref{ek:scheme_lagr}), one can construct new invariant schemes corresponding to
the representation~(\ref{derivHts}) of~$H^\prime(x)$. As an example we consider the
following one
\begin{equation} \label{ek:scheme_lagr_bottom_v2}
  \displaystyle
  \def\arraystretch{1.75}
  \large
  \begin{array}{c}
      \frac{1}{2}\left(
            (\hat{x}_s \check{x}_s)^{-\frac{1}{2}}
            + (\hat{x}_{\bar{s}} \check{x}_{\bar{s}})^{-\frac{1}{2}}
      \right)\!
      \left[
        \frac{x_s + x_{\bar{s}}}{2} \, x_{t\check{t}}
        - {\DSM}{H(x)}
      \right]
      + {\DSM}\left(
        (\hat{x}_s \check{x}_s)^{-1}
      \right)
      = 0,
      \\
      \tau_+ = \tau_-,
      \qquad
      h^s_+ = h^s_-.
  \end{array}
\end{equation}
Scheme~(\ref{ek:scheme_lagr_bottom_v2}) possesses
the conservation law of mass~(\ref{ek:mixed_derivs_cl})
and the conservation law of momentum~(which corresponds to the generator~$X_2$)
\[
    \DTM\left(
        x_t \frac{\hat{x}_s + \hat{x}_{\bar{s}}}{2}
    \right)
    +
    \DSM\left(
        2 (\hat{x}_s \check{x}_s)^{-\frac{1}{2}}
        - \frac{x_t x_t^+}{2}
        - H(x)
    \right) = 0
\]
with integrating factor
\[
    \frac{
        2 (\hat{x}_s \check{x}_s \hat{x}_{\bar{s}} \check{x}_{\bar{s}})^{\frac{1}{2}}
    }{
        (\hat{x}_s \check{x}_s)^{\frac{1}{2}}
        + (\hat{x}_{\bar{s}} \check{x}_{\bar{s}})^{\frac{1}{2}}
    }.
\]

Using (\ref{ek:contact_transform_lagr2}), (\ref{ek:state_eq_raw2}), one obtains
the following representation of the latter conservation law
%of the scheme (\ref{ek:scheme_lagr_bottom})
in Lagrangian mass coordinates
\[
    \displaystyle
    %\Lambda = \frac{u + \check{u}}{2}, \qquad
        \DTM\left(
            u \left(
                \frac{1}{\rho}
                + \frac{1}{\rho^-}
                - \frac{1}{2\sqrt{p}}
                - \frac{1}{2\sqrt{p^-}}
            \right)
    \right)
    + \DSM\left(
        2\sqrt{Q} - \frac{u u^+}{2} - H(x)
    \right) = 0,
\]
where $Q$ is given by the formula (\ref{ek:flux_Q}).

\subsubsection{Case $H^\prime(x) = \mathrm{const}$ (linear bottom).}

Here the function $H$ is $H(x) = C_1 x + C_2$,
and the scheme (\ref{ek:scheme_lagr_bottom}) becomes
\[
  \displaystyle
  \def\arraystretch{1.75}
  \large
  \begin{array}{c}
      x_{t\check{t}}
        + \frac{1}{h^s_-} \left(
        (\hat{x}_s \check{x}_s)^{-1}
        - (\hat{x}_{\bar{s}} \check{x}_{\bar{s}})^{-1}
      \right)
      - C_1
      = 0,
      \\
      \tau_+ = \tau_-,
      \qquad
      h^s_+ = h^s_-.
  \end{array}
\]

In case $C_1 = 0$ (a flat bottom), all the results for the latter scheme
coincide with the results of Section~\ref{ek:sec:delta_case_flat}.

In case $C_1 \neq 0$, the scheme admits the whole 6-parametric group of transformations~(\ref{ek:syms6}).
Using the difference analogue
\[
x = \tilde{x} + \frac{C_1}{2} t\hat{t}
\]
of transformation (\ref{ek:lin_to_flat}), one arrives at the flat bottom case, which was described above.

\section{Numerical results}

For comparison, we consider several schemes in Lagrangian mass coordinates.

\begin{enumerate}[label=(\alph*)]
    \item
    Invariant scheme~(\ref{ek:scheme_mass_coord2}),(\ref{ek:state_eq_raw2}),
    which was constructed in the previous sections.

    \item
    An explicit difference scheme, which is a simple difference approximation of
    the shallow water equations system.

    \item
    Samarskiy and Popov's completely conservative scheme for the gas dynamics equations~ \cite{bk:SamarskyPopov_book[1992]}, adapted for the shallow water equations.
    \item
    The completely conservative Yelenin--Krylov scheme for the two-layer shallow water
    equations~\cite{bk:YeleninKrylov[1982]}, adapted for the single-layer case.
    \item
    A completely conservative scheme proposed in the paper~\cite{bk:Korobitsyn_scheme[1989]} by V.~A.~Korobitsyn.
\end{enumerate}

All the schemes are considered on uniform orthogonal meshes in case of the flat bottom~($h(x) = 0$) only.
Also, unless specifically indicated, it is assumed that~$x_t = u$.
Consider schemes (b)---(e) in detail.

\paragraph{(b) An explicit scheme}

Consider the following explicit scheme, which approximates the equations (\ref{Lag1}), (\ref{Lag2})
and can be written in the divergent form as follows.
\begin{equation}\label{ek:naive}
\begin{array}{c}
    \frac{1}{\tau}\left(\frac{1}{\hat{\rho}} -  \frac{1}{\rho}\right) = \frac{u^+ - u}{h},
    \\
    \frac{\hat{u} - u}{\tau}  + \frac{1}{h} (\rho \hat{\rho} - \rho^- \hat{\rho}^-) = 0.
\end{array}
\end{equation}
These equations are the conservation laws of mass and momentum:
\[
\begin{array}{c}
    \DTP\left( \frac{1}{\rho} \right) - \DSP(u) = 0,
    \\
    \DTP(u) + \DSP(\rho^-\hat{\rho}^-) = 0.
\end{array}
\]
The center of mass conservation law is
\[
  \DTP(\check{t} u - x) + \DSP(t \rho\hat{\rho}) = 0.
\]
The energy conservation law for the scheme~(\ref{ek:naive}) does not hold:
\[
\def\arraystretch{1.75}
\begin{array}{c}
   \DTP\left( \frac{u^2}{2} + \rho \right)
   + \DSP\left( u \rho^-\hat{\rho}^- \right) = \\
   = u \left( \DTP(u) + \DSP(\rho^-\hat{\rho}^-) \right)
    - \rho \hat{\rho} \left( \DTP\left( \frac{1}{\rho} \right) - \DSP(u) \right)
    + \frac{1}{2}u_t^2\,\tau.
\end{array}
\]

\paragraph{(c) The Samarskiy-Popov scheme}

An implicit completely conservative schemes
by Samarskiy--Popov~\cite{bk:SamarskyPopov_book[1992]} for the
one-dimensional gas dynamics equations can be modified
for the case of the shallow water as follows.
\begin{equation} \label{ek:SamPopScheme}
        \def\arraystretch{2}
        \scalebox{1.25}{$
        \begin{array}{c}
        u_t + \hat{p}_{\bar{s}} = 0,
        \qquad
        \frac{1}{\rho} = x_s,
        \\
        x_t = \frac{\hat{u} + u}{2},
        \qquad
        p = A \rho^2.
        \end{array}
        $}
\end{equation}
Scheme~(\ref{ek:SamPopScheme}) is invariant.

The original Samarskiy--Popov scheme for the one-dimensional gas dynamics equations
also includes a conservation law of energy:
\begin{equation}\label{Samarsky_energy}
    \scalebox{1.25}{$
        \DTP\left(\varepsilon + \frac{u_+^2}{2} \right)
            + \DSP\left(\frac{(u + \hat{u})(p + \hat{p})}{4}\right) = 0,
    $}
\end{equation}
where the internal energy of the medium~$\varepsilon$ is determined by the equation
\begin{equation}\label{Samarsky_energy_nondiv}
    \varepsilon_t = -\hat{p} \left(\frac{1}{\rho}\right)_t,
\end{equation}
relating the change in internal energy with the work of pressure forces.

In the case of shallow water, in contrast to the original scheme for the gas dynamics equations,
the energy equation~(\ref{Samarsky_energy}) is no longer a part of the scheme
and it does not directly follow from~(\ref{ek:SamPopScheme}).
One can rewrite equation~(\ref{Samarsky_energy}), taking into account~(\ref{Samarsky_energy_nondiv}),
as follows:
\begin{equation}\label{SamPop_no_energy_cl}
  \frac{\hat{u}_+ + u_+}{2} (u^+_t + \hat{p}_s)
  - \hat{p} \left( \left(\frac{1}{\rho}\right)_t - \frac{1}{2}\left(\hat{u} + u \right)_s \right)
  - \frac{1}{4}( (u + \hat{u}) p_t )_s \tau = 0.
\end{equation}
Then, one sees that equation~(\ref{Samarsky_energy}) does not hold on the solutions of system~(\ref{ek:SamPopScheme}),
and, strictly speaking, it is not a conservation law of the system.
Nevertheless, we are going to use equations~(\ref{Samarsky_energy}) and~(\ref{Samarsky_energy_nondiv})
of the original Samarskiy--Popov scheme
in numerical computations in order
to control the conservation law of energy on the solutions of the system.
%%%%%%%%%%%%%%%%%%%%%%%%%%%%%%%%%%%%%%%%%%%%%%%
\paragraph{(d) The Yelenin--Krylov scheme}

This is a completely conservative scheme for a system of equations of
the two-layer shallow water obtained in~\cite{bk:YeleninKrylov[1982]}.
It can be reduced to the system of equations of single-layer shallow water
by assuming all the parameters of the both layers to be equal.
Then, after some simplifications, one arrives at the following scheme:
\begin{equation}\label{ek:Yelenin3}
    \def\arraystretch{1.75}
    \begin{array}{c}
        \DTP\left( \frac{1}{\rho} \right) - \DSP(u^{(0.5)}) = 0,
        \\
        \DTP(u) + \DSP\left(\underset{-s}{S}\left(
                P^{(0.5)}
                + (\rho^{-1})^{(0.5)}
                (P^\prime)^{(0.5)}
        \right)\right) = 0,
        \\
    \end{array}
\end{equation}
where
\[
    f^{(0.5)} = \frac{\hat{f} + f}{2}
\quad
\mbox{and}
\quad
 %   F^{(0.5)} = P^{(0.5)} - (\rho^{-1})^{(0.5)} (P^\prime)^{(0.5)},
  %  \qquad
    P^\prime = \frac{\hat{P} - P}{\hat{\rho}^{-1} - \rho^{-1}},
\]
the function $P$ is given by a state equation and
in the case of the single-layer shallow water it should
be equal to~$\rho^2$ in the continuous limit.

The scheme possesses the conservation law of energy
\[
\DTP\left(
            \frac{u^2 + (u^+)^2}{4} + \frac{P}{\rho}
        \right)
        + \DSP\left(
           u^{(0.5)} \left(\frac{F + F^-}{2}\right)^{(0.5)}
        \right) = 0,
        %x_t = u,
\]
where
\[
    F^{(0.5)} = P^{(0.5)} - (\rho^{-1})^{(0.5)} (P^\prime)^{(0.5)}.
\]

In the given form, scheme~(\ref{ek:Yelenin3}) is a three-level one.
In order to write it on two time levels, one sets
the state equation that defines the function $P$ as follows:
\begin{equation}\label{ek:Yelenin_state}
  \hat{P} - P = 2 (\hat{\rho}^{-1} - \rho^{-1}) \rho^3.
\end{equation}
The latter expression corresponds to the differential
\[
\frac{dG}{d\rho} = \frac{d}{d\rho}(\rho^2) = 2 \rho.
\]
Taking into account equation~(\ref{ek:Yelenin_state}),
one can write a modified Yelenin--Krylov scheme as follows
\begin{equation}
    \label{ek:Yelenin}
    \def\arraystretch{1.75}
    \begin{array}{c}
        \DTP\left( \frac{1}{\rho} \right) - \DSP(u^{(0.5)}) = 0,
        \\
        \DTP(u) + \DSM\left(
                P^{(0.5)}
                + (\rho^{-1})^{(0.5)}
                (2\rho^3)^{(0.5)}
        \right) = 0,
        \\
        \hat{P} - P = 2 (\hat{\rho}^{-1} - \rho^{-1}) \rho^3,
        \\
        \DTP\left(
            \frac{1}{4} (u^2 + (u^+)^2) + \frac{1}{\rho}P
        \right) +
        \\
        \qquad + \frac{1}{2}\DSP\left(
           h^s u^{(0.5)} \DSM\left(
                P^{(0.5)} + (\rho^{-1})^{(0.5)} (2\rho^3)^{(0.5)}
           \right)
        \right) = 0.
        %x_t = u,
    \end{array}
\end{equation}
%%%%%%%%%%%%%%%%%%%%%%%%%%%%%%%%%%%%%%%%%%
\paragraph{(e) The Korobitsyn scheme}
This scheme was proposed in the paper~\cite{bk:Korobitsyn_scheme[1989]},
and it is a modification of the completely conservative Samarskiy-Popov
scheme~\cite{bk:SamarskyPopov_book[1992]}.
For the shallow water equations, according to~(\ref{ek:sw_stetEq2}),
one can rewrite the scheme in the following form
\begin{equation} \label{ek:KorobScheme}
    \def\arraystretch{1.75}
  \begin{array}{c}
    u_t + p_{\bar{s}} = 0,
    \\
    x_t = 0.5 (u + \hat{u}),
    \\
    \left( \frac{1}{\rho} \right)_t = 0.5 (u + \hat{u})_{\bar{s}},
    \\
    \hat{p} = \left( 0.5 q (\rho_- + \rho) + (1- q) \rho \right)
                \left( \rho + 0.25 q \tau^2 u_t^2 \right),
  \end{array}
\end{equation}
where $0 \leqslant q \leqslant 1$ is a parameter.
Setting $q=0$, one gets one of the Samarskiy--Popov schemes~\cite{bk:SamarskyPopov[1970]}.
If $q=1$, scheme~(\ref{ek:KorobScheme})
becomes ``thermodynamically consistent''~(see.~~\cite{bk:Korobitsyn_scheme[1989]}),
i.~e., it possesses additional conservation laws if the one-dimensional flow
of polytropic gas with an adiabatic exponent~$\gamma=3$ is considered.
In the shallow water case, the condition of thermodynamic consistency does not have
of clear physical meaning, but further we put $q = 1$, as it was done in~\cite{bk:Korobitsyn_scheme[1989]}.
In this case, the scheme can be represented in an explicit form.
Using the relations given in~\cite{bk:Korobitsyn_scheme[1989]},
one can also derive the following expression for the conservation law of energy
\[
\def\arraystretch{1.75}
\begin{array}{c}
  \left(\frac{u^2 + u_+^2}{2}\right)_t
  + \Big(
    \frac{\hat{u} + u}{8} \Big[
        (\rho + \rho_-)(\rho + 0.25 q \tau^2 u_t^2)
        \\
        \qquad\qquad
        + (\rho_- + \rho_{--})(\rho_- + 0.25 q \tau^2 (u^-_t)^2)
    \Big]
    \Big)_s = 0.
\end{array}
\]

%%%%%%%%%%%%%%%%%%%%%%%%%%%%%%%%%%%%%%%%%%%%%%%%%%%%%%%%%%%%%%%%%%%
\bigskip

In order to reduce the schemes' dispersion,
we use an artificial linear-quadratic viscosity~\cite{bk:SamarskyPopov_book[1992]}
of the form
\[
  \omega = -\nu \rho u_s + 0.5 (1 + \gamma) \rho \frac{\kappa h^s}{\pi} u_s^2,
\]
where $\nu$ is a linear viscosity coefficient,
$\kappa$ is a quadratic viscosity coefficient,
and $\gamma$ --- adiabatic exponent, which equals~$2$
in the case of shallow water.
The use of artificial viscosity means that in all the calculations
the value of pressure~$p$ near strong discontinuities of solutions
is replaced by
\[
  q = p + \omega.
\]

\medskip

Further, four test problems are considered
(detailed description of three of them one can find in~\cite{bk:SamarskyPopov_book[1992]}).
In all the tasks the mesh space steps are given as~$h^s=0.02$ and~$\tau = 0.025 h^s$.
Viscosity coefficients $\nu$ and $\kappa$ were chosen empirically
so that solutions of the schemes to be close to the known exact solutions.
For the modified Yelenin-Krylov scheme~(\ref{ek:Yelenin}) artificial viscosity is not used,
as it turned out, in this case it markedly worsens the results.
For the explicit scheme we put~$\nu=0.005$, and, for scheme~(\ref{ek:KorobScheme})
and for the modified Samarskiy--Popov scheme~(\ref{ek:SamPopScheme}), we put~$\nu=0.001$.
In all cases, except for the scheme~(\ref{ek:Yelenin}), the coefficient of quadratic viscosity is~$\kappa=4.5$.
The total substance mass~$S$ for the first three tests is taken~$S = 3$,
and for the fourth test~$S=4$.
The initial height of the fluid column~$\rho_0$ in all cases is~$1$, and the pressure~$p_0$
is calculated by the equations of state.
Calculations for all the implicit schemes were carried out by iterative methods
with the accuracy requirement~$\varepsilon = 0.001$.

\begin{figure}[ht]
\begin{minipage}[c]{\linewidth}\centering
\includegraphics[width=0.25\linewidth]{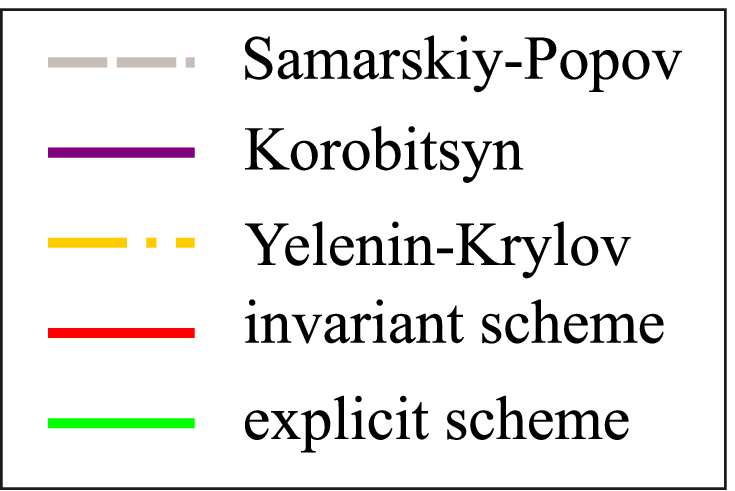}
\vspace{0.2cm}
\end{minipage}
\caption{Legend for Fig.~\ref{appB:pic1}--\ref{appB:pic4}}
\label{appB:pic0}
\end{figure}

Descriptions of the test problems and the numerical results are presented below.

Notice that in Figures~\ref{appB:pic1}--\ref{appB:pic4}
the legend which is depicted in Figure~\ref{appB:pic0} is used.

\begin{itemize}
  \item
  \emph{Test 1. Rarefaction wave}.
  This is the problem of a withdrawing piston, which, under the action of
  external forces, moves outside of the medium producing a rarefaction wave.
  The numerical results, as well as the values of the conservation laws on solutions,
  for all the schemes at the moment~$t=0.55$ are shown in the Figure~\ref{appB:pic1}.
  The height~$\rho$ of the fluid column above the flat bottom at the chosen moment of time
  is shown in Figure~\ref{appB:pic1}~a).
  The values of the conservation laws of energy, mass and momentum on
  the solution Figure~\ref{appB:pic1}~a) are depicted in
  Figures~\ref{appB:pic1}~b)---~\ref{appB:pic1}~d) appropriately.

  \item
  \emph{Test 2. Compression wave}.
  Here the piston moves into the medium at a constant speed~$U_0 = 0.5$,
  producing a shock wave.
  The numerical results at the moment~$t=0.6$ are given in Figure~\ref{appB:pic2}
  by analogy with Figure~\ref{appB:pic1}.

  \item
  \emph{Test 3. Unidirectional compression waves}.
  At the initial moment of time under the action of a piston moving at a constant speed $U_0 = 0.8$
  to the right, there is a compression wave.
  At time~$t_1 = 0.25$, the piston begins to accelerate until the moment~$t_2 = 0.5$,
  when its speed reaches the value~$U_1 = 2 U_0 = 1.6$. After that the piston moves
  at a constant speed $U_1$. The piston moves according to the following law
  \[
    \displaystyle
    U(t) = \begin{cases}
             U_0, \qquad t \leqslant t_1,
             \\
             U_0 + (U_1 - U_0) \sin^2 \left(\frac{\pi}{2} \frac{t - t_1}{t_2 - t_1}\right),
             \quad t_1 \leqslant t \leqslant t_2,
             \\
             U_1, \qquad t \geqslant t_2.
           \end{cases}
  \]
  The numerical results at~$t=0.74$ are given in Figure~\ref{appB:pic3}.

  \item
  \emph{Test 4. Collision of two opposing compression waves}.
  In this test, two compression waves move towards each other at speeds~$U_1 = 0.5$ and~$U_2 = -0.5$.
  The numerical results at~$t=1.0$ are given in Figure~\ref{appB:pic4}.
\end{itemize}

\begin{remark}
One can find some exact smooth solutions and generalizations of the scheme
and the corresponding numerical computations in the paper~\cite{bk:EK_preprint[2019]}.
\end{remark}

Comparing the results of calculations according to five schemes,
we arrive at the following conclusions.
\begin{enumerate}
    \item
    All the considered schemes,
    with proper selection of artificial viscosity,
    show numerical results close to the known exact solutions.

    \item
    The solutions of the modified Yelenin--Krylov scheme~(\ref{ek:Yelenin})
    change with dispersal peaks near discontinuities,
    while the artificial viscosity value does not significantly affect that peaks.
\end{enumerate}

\newpage
\begin{landscape}

\vspace*{\fill}

\begin{figure}[ht]
      \begin{minipage}[b][][b]{0.24\linewidth}\centering
        \includegraphics[width=\linewidth]{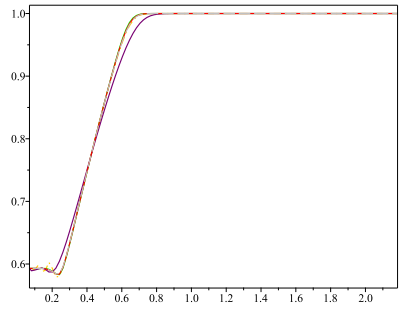} \\a) test
      \end{minipage}
      \hfill
      \begin{minipage}[b][][b]{0.24\linewidth}\centering
        \includegraphics[width=\linewidth]{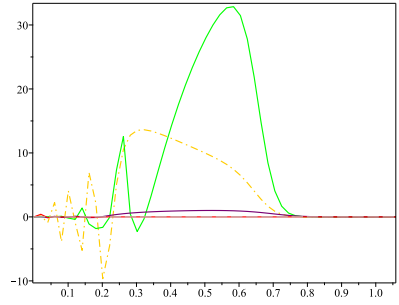}\\b) energy
      \end{minipage}
      \begin{minipage}[b][][b]{0.24\linewidth}\centering
        \includegraphics[width=\linewidth]{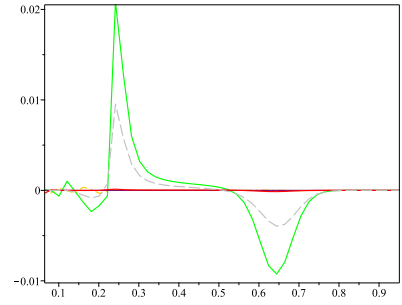} \\c) mass
      \end{minipage}
      \hfill
      \begin{minipage}[b][][b]{0.24\linewidth}\centering
        \includegraphics[width=\linewidth]{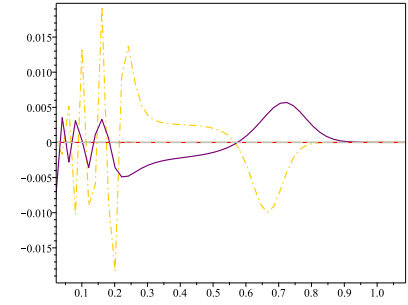}\\d) momentum
      \end{minipage}
      \caption{Test 1 numerical results}
      \label{appB:pic1}
  \end{figure}

  \vspace*{\fill}

  \begin{figure}[ht]
      \begin{minipage}[b][][b]{0.24\linewidth}\centering
        \includegraphics[width=\linewidth]{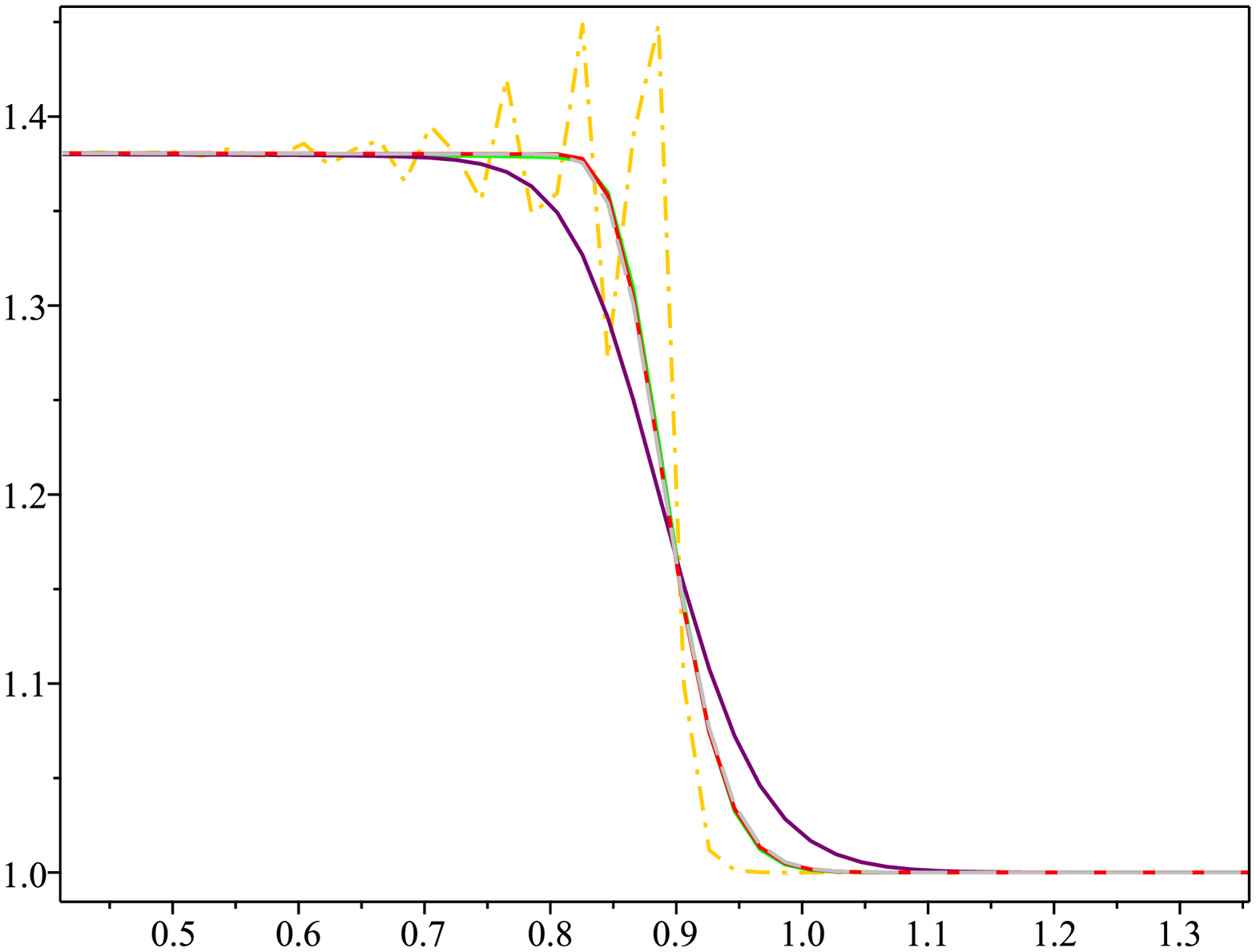} \\a) test
      \end{minipage}
      \hfill
      \begin{minipage}[b][][b]{0.24\linewidth}\centering
        \includegraphics[width=\linewidth]{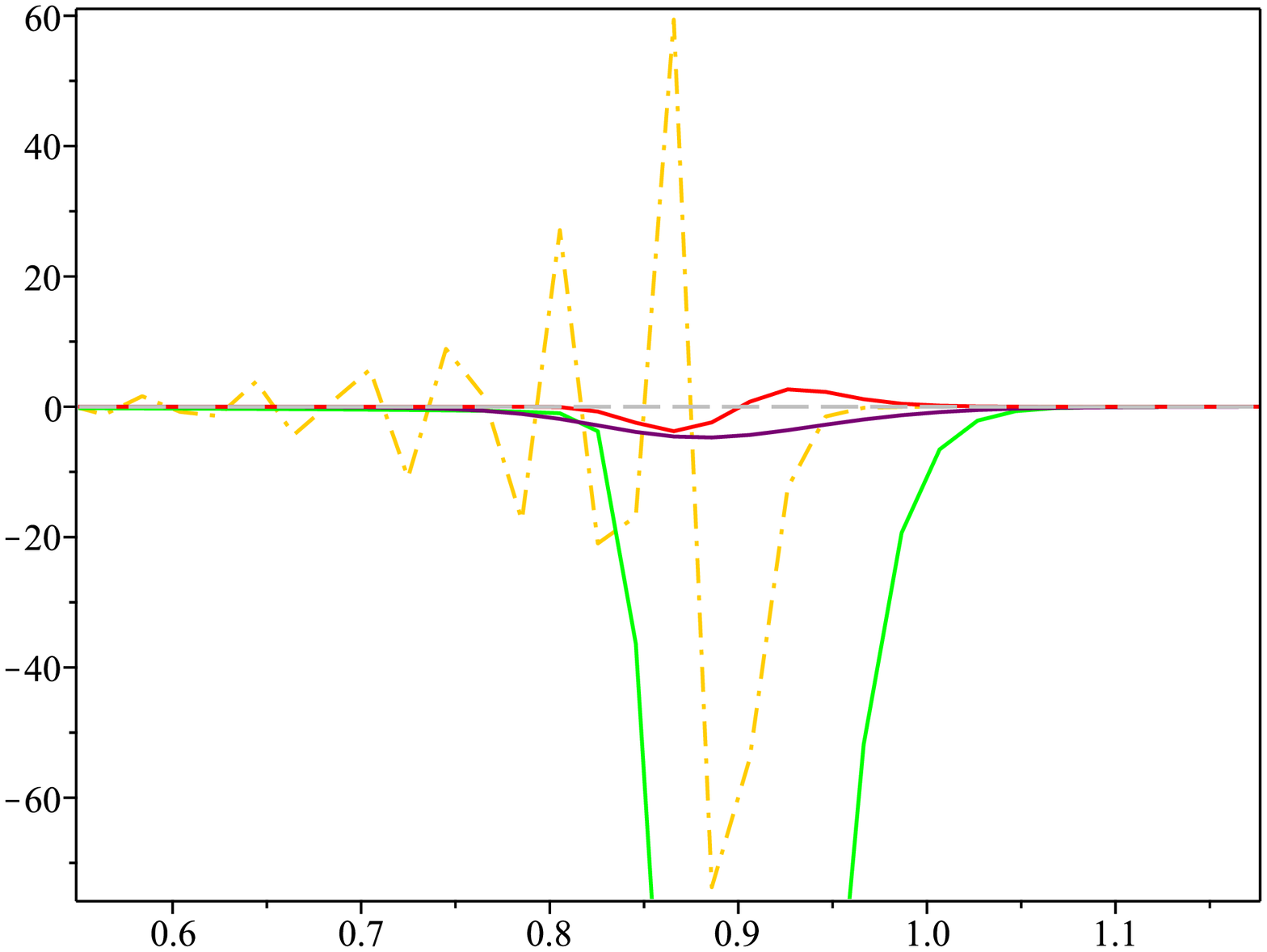}\\b) energy
      \end{minipage}
      \begin{minipage}[b][][b]{0.24\linewidth}\centering
        \includegraphics[width=\linewidth]{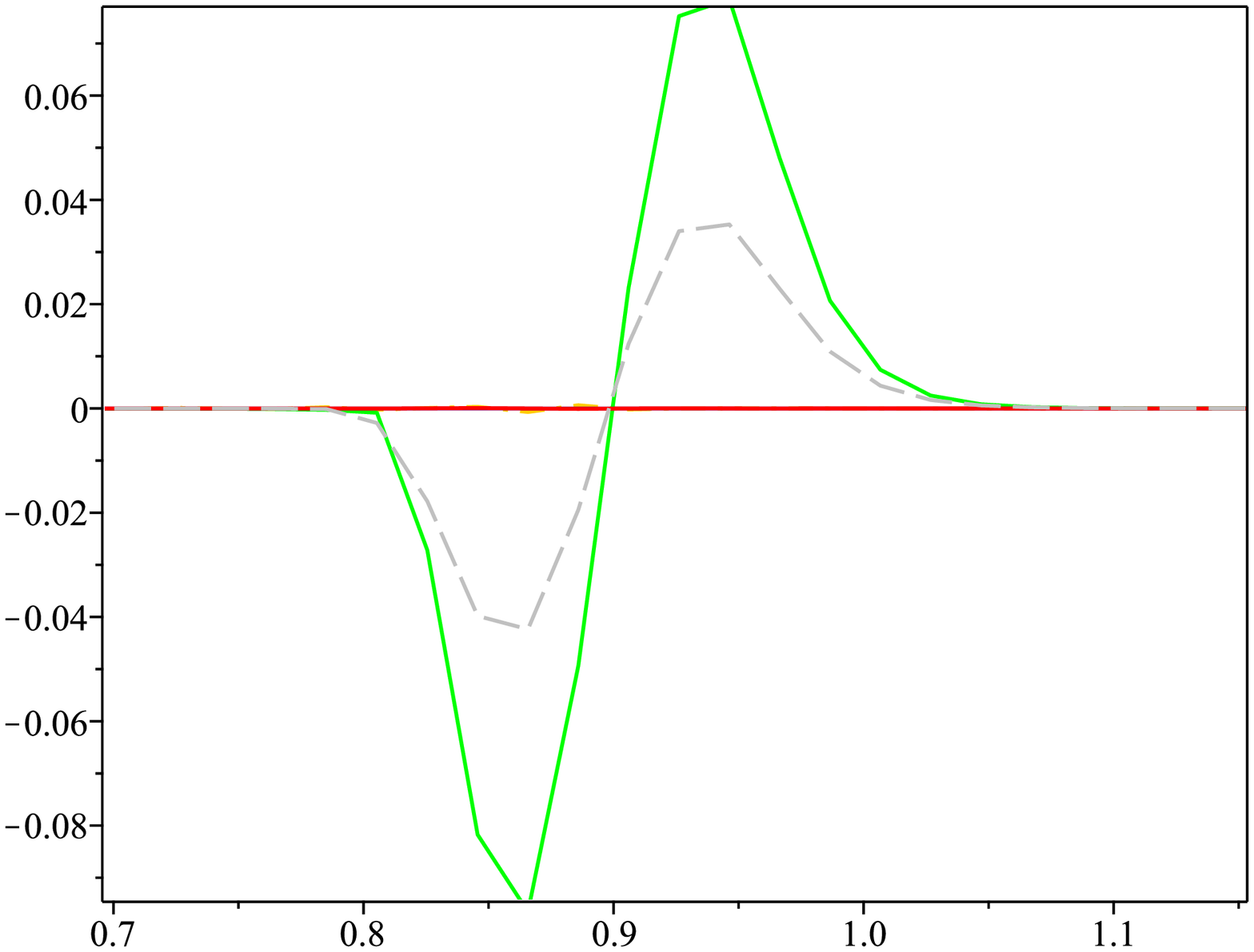} \\c) mass
      \end{minipage}
      \hfill
      \begin{minipage}[b][][b]{0.24\linewidth}\centering
        \includegraphics[width=\linewidth]{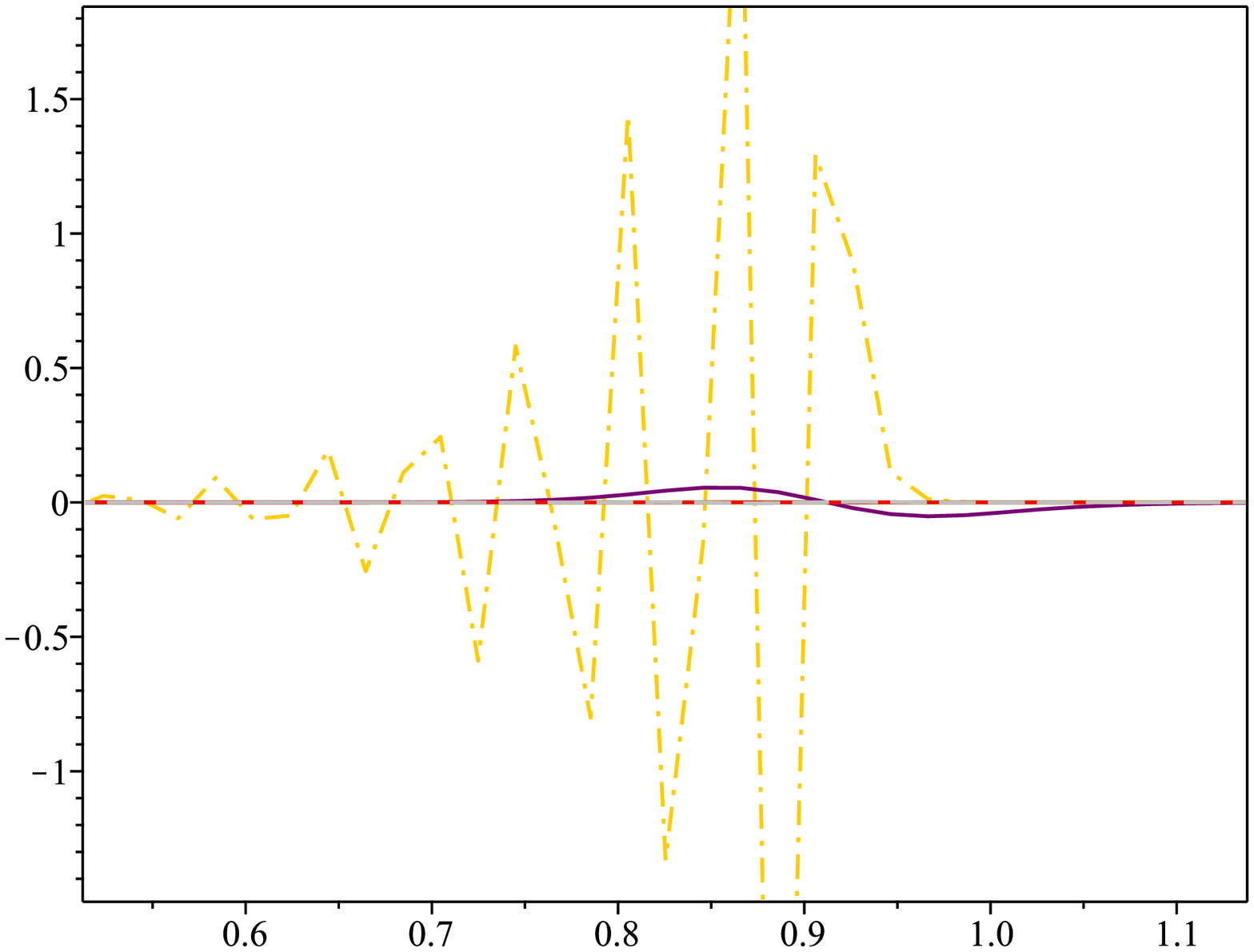}\\d) momentum
      \end{minipage}
      \caption{Test 2 numerical results}
      \label{appB:pic2}
  \end{figure}

  \vspace*{\fill}

  \end{landscape}

    \newpage

  \begin{landscape}

    \vspace*{\fill}

  \begin{figure}[ht]
      \begin{minipage}[b][][b]{0.24\linewidth}\centering
        \includegraphics[width=\linewidth]{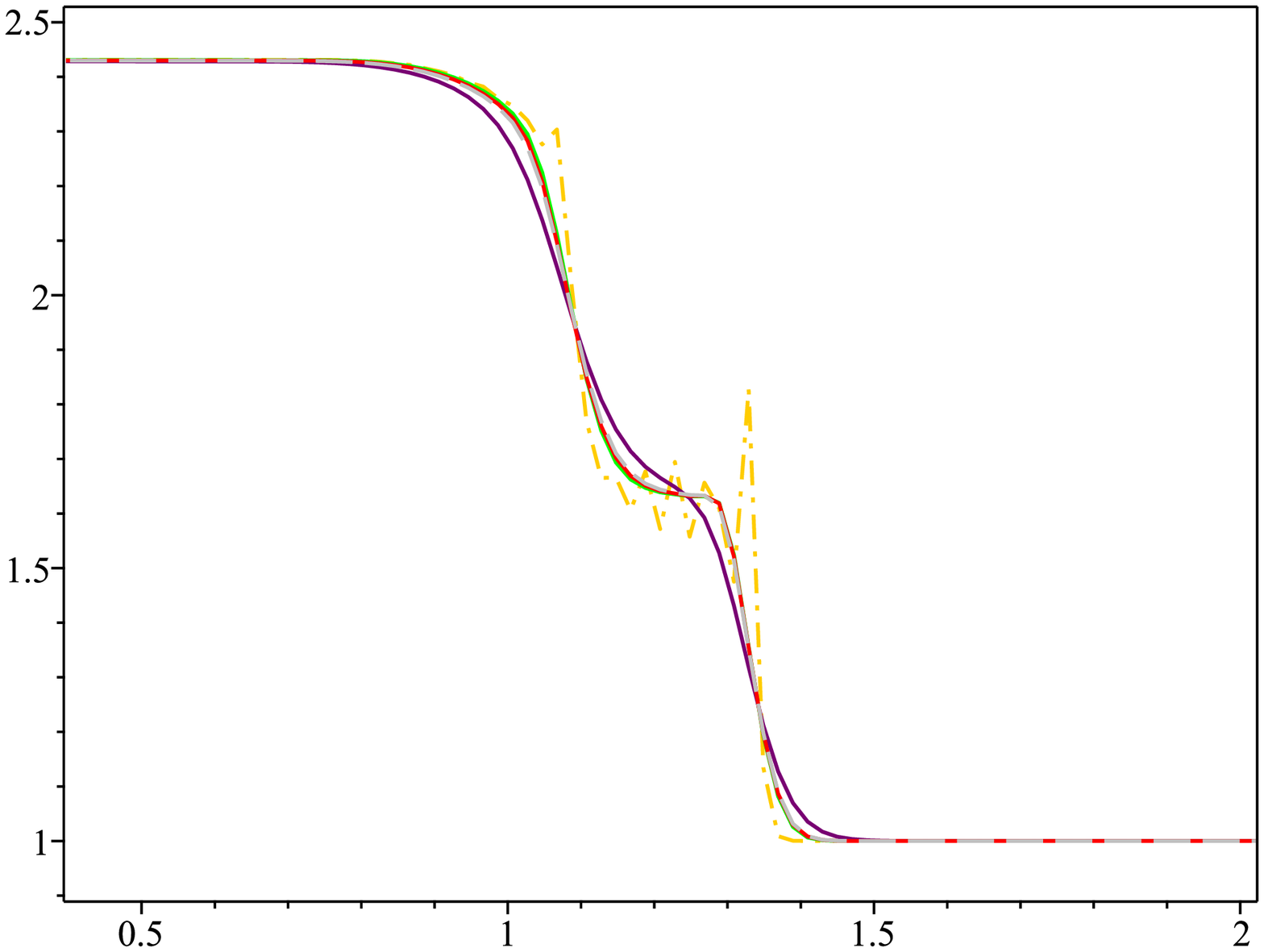} \\a) test
      \end{minipage}
      \hfill
      \begin{minipage}[b][][b]{0.24\linewidth}\centering
        \includegraphics[width=\linewidth]{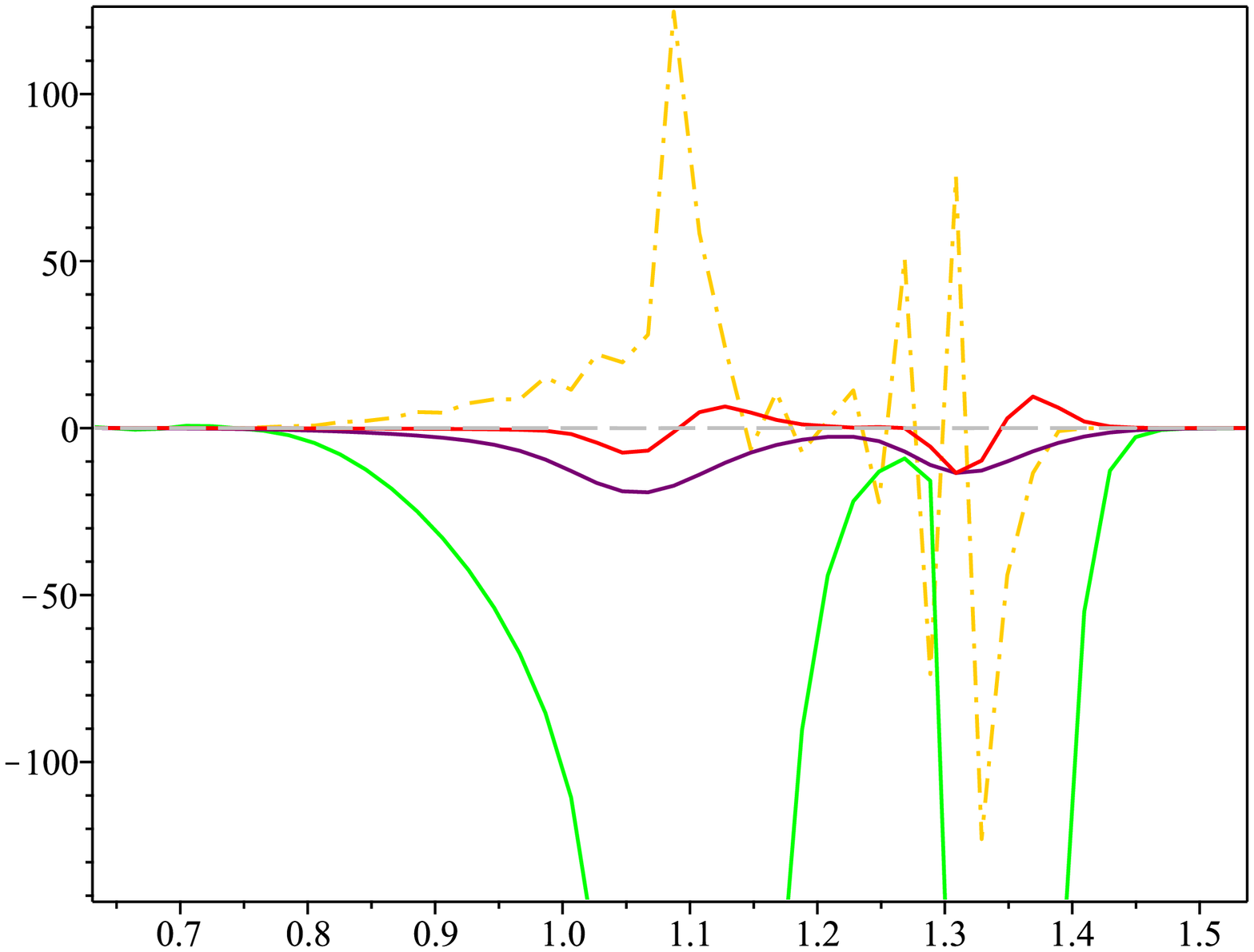}\\b) energy
      \end{minipage}
      \begin{minipage}[b][][b]{0.24\linewidth}\centering
        \includegraphics[width=\linewidth]{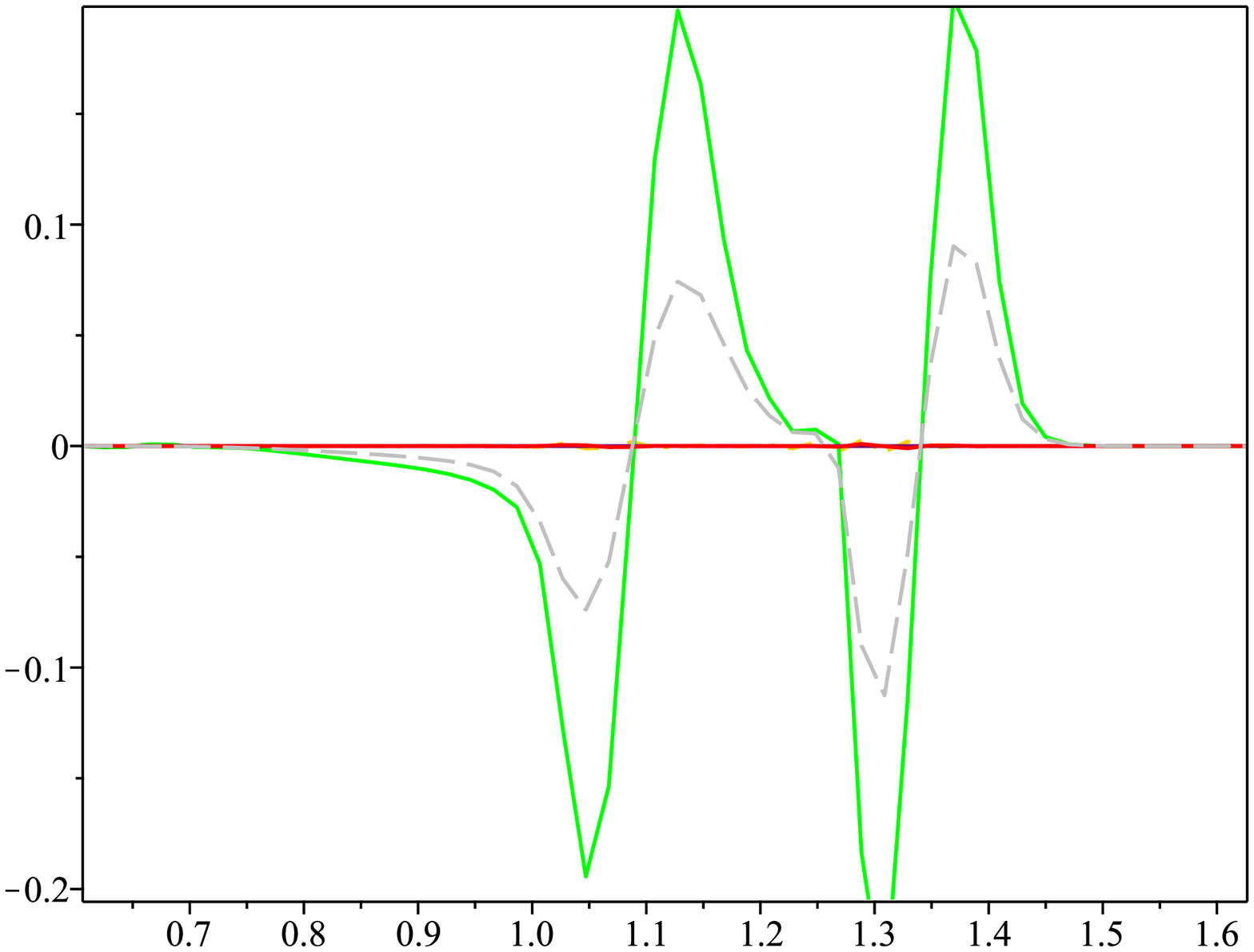} \\c) mass
      \end{minipage}
      \hfill
      \begin{minipage}[b][][b]{0.24\linewidth}\centering
        \includegraphics[width=\linewidth]{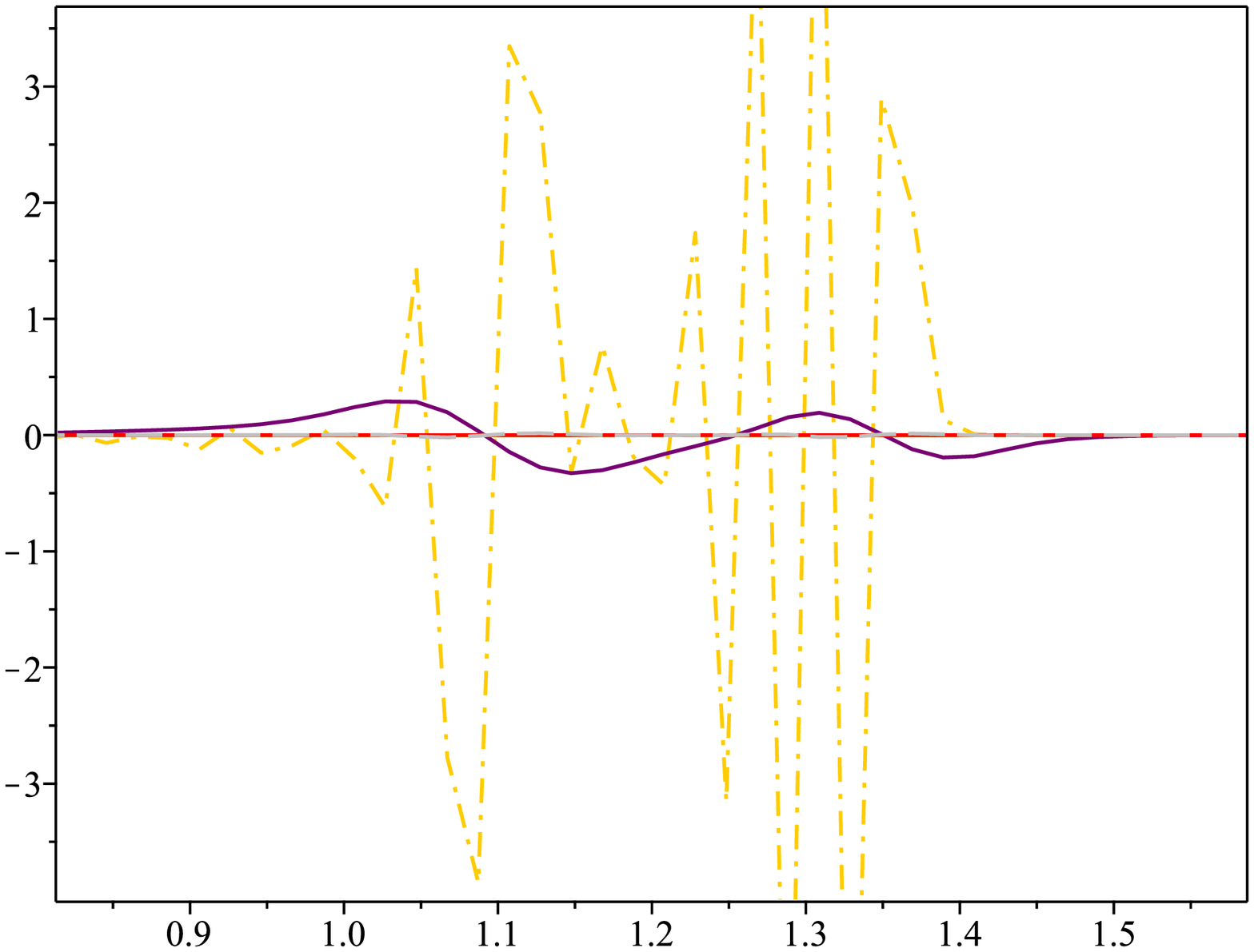}\\d) momentum
      \end{minipage}
      \caption{Test 3 numerical results}
      \label{appB:pic3}
  \end{figure}

  \vspace*{\fill}

  \begin{figure}[ht]
      \begin{minipage}[b][][b]{0.24\linewidth}\centering
        \includegraphics[width=\linewidth]{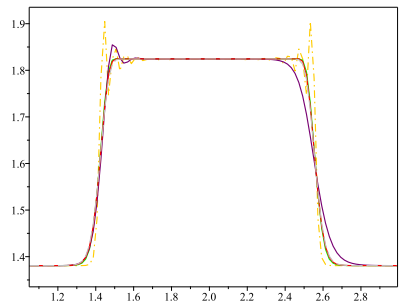} \\a) test
      \end{minipage}
      \hfill
      \begin{minipage}[b][][b]{0.24\linewidth}\centering
        \includegraphics[width=\linewidth]{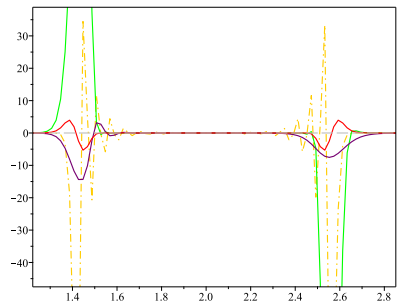}\\b) energy
      \end{minipage}
      \begin{minipage}[b][][b]{0.24\linewidth}\centering
        \includegraphics[width=\linewidth]{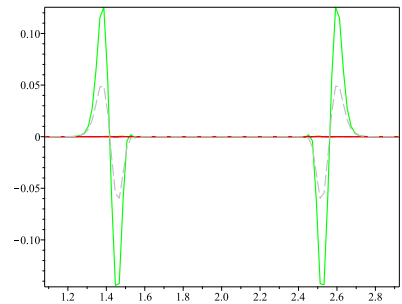} \\c) mass
      \end{minipage}
      \hfill
      \begin{minipage}[b][][b]{0.24\linewidth}\centering
        \includegraphics[width=\linewidth]{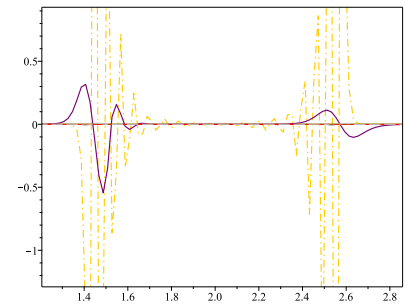}\\d) momentum
      \end{minipage}
      \caption{Test 4 numerical results}
      \label{appB:pic4}
  \end{figure}

    \vspace*{\fill}

\end{landscape}

% это продолжение перечисления, которое разорвано, чтобы влезли графики
\begin{enumerate}[resume]
   \item
    The modified Korobitsyn's scheme~(\ref{ek:KorobScheme}) turns out to be
    extremely sensitive to the artificial viscosity parameters,
    which makes big difference from scheme~(\ref{ek:Yelenin}).
    In contrast to the implicit modified Samarskiy--Popov's scheme~(\ref{ek:SamPopScheme})
    the explicit scheme~(\ref{ek:KorobScheme}) shows worse results according
    to the conservation of energy and momentum, but it better conserves mass.

    \item
    In all the cases, the invariant scheme~(\ref{ek:scheme_mass_coord2}), (\ref{ek:state_eq_raw2})
    results appear among the best results on the conservation laws on solutions.
\end{enumerate}

\section{Conclusion}

The one-dimensional shallow water equations in Eulerian and Lagrangian coordinates
were considered in the present work.

It was shown the relationship between symmetries and conservation laws in Lagrangian (potential) coordinates and symmetries and conservation laws in physical variables.

For equations in Lagrangian coordinates with a flat bottom an invariant difference scheme is
constructed which possesses all the difference analogues of the differential conservation laws: mass, momentum, energy,
the law of center of mass motion.

The resulting conservative scheme is three-layers scheme in time.
By non-point transformations of variables such a scheme can be reduced to an invariant scheme in mass
Lagrangian variables possessing the same set of conservation laws.
By choosing a special form of ``equation of state''~(the relationship of pressure and height of a liquid column)
it is possible to obtain a two-level scheme of shallow water equations in physical variables.
Generalization of the scheme to the case of an arbitrary bottom faces certain difficulties,
leading to that it is possible to construct either an invariant scheme with conservation of mass and momentum, or an invariant scheme with conservation of mass and energy.

For a conservative difference scheme for a flat bottom
some invariant solutions are constructed.
It is shown that the invariant scheme admits
reduction on subgroups as well as the original system of differential equations.

Invariant conservative difference scheme for the case of a flat bottom
tested numerically in comparison with other known schemes adapted
for the case of shallow water in one-dimensional approximation.
The numerical tests indicated a good accuracy of calculations and the validation
of conservation laws on solutions with big gradients.

\section*{Acknowledgements}
The research  was supported by Russian Science Foundation Grant No
18-11-00238 `Hydrodynamics-type equations: symmetries, conservation
laws, invariant difference schemes'.
E.I.K. also acknowledges Suranaree University of Technology for
Full-time Master Researcher Fellowship~(15/2561).

%plainnat
\bibliographystyle{elsarticle-num-names}
%\bibliographystyle{elsarticle-num-names}
%\inputencoding{utf8}
%\bibliography{references}

\begin{thebibliography}{67}
\expandafter\ifx\csname natexlab\endcsname\relax\def\natexlab#1{#1}\fi
\providecommand{\url}[1]{\texttt{#1}}
\providecommand{\href}[2]{#2}
\providecommand{\path}[1]{#1}
\providecommand{\DOIprefix}{doi:}
\providecommand{\ArXivprefix}{arXiv:}
\providecommand{\URLprefix}{URL: }
\providecommand{\Pubmedprefix}{pmid:}
\providecommand{\doi}[1]{\href{http://dx.doi.org/#1}{\path{#1}}}
\providecommand{\Pubmed}[1]{\href{pmid:#1}{\path{#1}}}
\providecommand{\bibinfo}[2]{#2}
\ifx\xfnm\relax \def\xfnm[#1]{\unskip,\space#1}\fi
%Type = Book
\bibitem[{Whitham(1974)}]{bk:Whitham[1974]}
\bibinfo{author}{G.~B. Whitham}, \bibinfo{title}{Linear and Nonlinear Waves},
  \bibinfo{publisher}{Wiley}, \bibinfo{address}{New York},
  \bibinfo{year}{1974}.
%Type = Book
\bibitem[{Ovsiannikov(2003)}]{bk:Ovsyannikov[2003]}
\bibinfo{author}{L.~V. Ovsiannikov}, \bibinfo{title}{Lectures on the gas
  dynamics equations}, \bibinfo{publisher}{Institute of Computer Studies},
  \bibinfo{address}{Moscow--Izhevsk}, \bibinfo{year}{2003}. \bibinfo{note}{In
  Russian}.
%Type = Book
\bibitem[{Petrosyan(2014)}]{bk:PetrosyanBook[2010]}
\bibinfo{author}{A.~S. Petrosyan}, \bibinfo{title}{Additional chapters of heavy
  fluid hydrodynamics with a free boundary}, \bibinfo{publisher}{Space Research
  Institute of the Russian Academy of Sciences}, \bibinfo{address}{Moscow},
  \bibinfo{year}{2014}. \bibinfo{note}{In Russian}.
%Type = Book
\bibitem[{Vallis(2006)}]{bk:Vallis[2006]}
\bibinfo{author}{G.~K. Vallis}, \bibinfo{title}{Atmospheric and Oceanic Fluid
  Dynamics: Fundamentals and Large-scale Circulation},
  \bibinfo{publisher}{Cambridge University Press},
  \bibinfo{address}{Cambridge}, \bibinfo{year}{2006}.
%Type = Article
\bibitem[{Bernetti et~al.(2008)Bernetti, Titarev, and Toro}]{bk:Bernetti[2008]}
\bibinfo{author}{R.~Bernetti}, \bibinfo{author}{V.~A. Titarev},
  \bibinfo{author}{E.~F. Toro},
\newblock \bibinfo{title}{Exact solution of the riemann problem for the shallow
  water equations with discontinuous bottom geometry},
\newblock \bibinfo{journal}{Journal of Computational Physics}
  \bibinfo{volume}{227} (\bibinfo{year}{2008}) \bibinfo{pages}{3212 -- 3243}.
  \DOIprefix\doi{https://doi.org/10.1016/j.jcp.2007.11.033}.
%Type = Article
\bibitem[{Han et~al.(2012)Han, Hantke, and Warnecke}]{bk:HanHantke[2012]}
\bibinfo{author}{E.~E. Han}, \bibinfo{author}{M.~Hantke},
  \bibinfo{author}{G.~Warnecke},
\newblock \bibinfo{title}{Exact riemann solutions to compressible {E}uler
  equations in ducts with discontinuous cross-section},
\newblock \bibinfo{journal}{Journal of Hyperbolic Differential Equations}
  \bibinfo{volume}{09} (\bibinfo{year}{2012}) \bibinfo{pages}{403--449}.
  \DOIprefix\doi{10.1142/S0219891612500130}.
%Type = Book
\bibitem[{Kulikovskii et~al.(2001)Kulikovskii, Pogorelov, and
  Semenov}]{bk:KulikovskyPogorelovSemenov}
\bibinfo{author}{A.~G. Kulikovskii}, \bibinfo{author}{N.~V. Pogorelov},
  \bibinfo{author}{A.~Y. Semenov}, \bibinfo{title}{Mathematical aspects of
  numerical solution of hyperbolic systems}, \bibinfo{publisher}{Chapman \&
  Hall/CRC Monographs and Surveys in Pure and Applied Mathematics},
  \bibinfo{address}{Boca Raton}, \bibinfo{year}{2001}.
%Type = Book
\bibitem[{Vreugdenhil(1994)}]{bk:Vreugdenhil[1994]}
\bibinfo{author}{C.~B. Vreugdenhil}, \bibinfo{title}{Numerical Methods for
  Shallow-Water Flow}, \bibinfo{publisher}{Springer Science+Business Media},
  \bibinfo{address}{Dordrecht}, \bibinfo{year}{1994}.
%Type = Book
\bibitem[{Tan(2012)}]{bk:TanWeiyan}
\bibinfo{author}{W.~Y. Tan}, \bibinfo{title}{Shallow Water Hydrodynamics:
  Mathematical Theory and Numerical Solution for a Two-dimensional System of
  Shallow Water Equations}, \bibinfo{publisher}{Water \& Power Press},
  \bibinfo{address}{Beijing}, \bibinfo{year}{2012}.
%Type = Book
\bibitem[{Abbasov(2018)}]{bk:Abbasov}
\bibinfo{author}{I.~B. Abbasov}, \bibinfo{title}{3D Modeling of Nonlinear Wave
  Phenomena on Shallow Water Surfaces}, \bibinfo{publisher}{Wiley},
  \bibinfo{address}{Hoboken}, \bibinfo{year}{2018}.
%Type = Article
\bibitem[{Yelenin and Krylov(1982)}]{bk:YeleninKrylov[1982]}
\bibinfo{author}{G.~G. Yelenin}, \bibinfo{author}{V.~V. Krylov},
\newblock \bibinfo{title}{A completely conservative difference scheme for
  equations of two-layered ``shallow water'' in {L}agrange coordinates},
\newblock \bibinfo{journal}{Differ. Uravn.} \bibinfo{volume}{18}
  (\bibinfo{year}{1982}) \bibinfo{pages}{1190–--1196}.
%Type = Article
\bibitem[{Bihlo and Popovych(2012)}]{bk:Bihlo_numeric[2012]}
\bibinfo{author}{A.~Bihlo}, \bibinfo{author}{R.~Popovych},
\newblock \bibinfo{title}{Invariant discretization schemes for the
  shallow-water equations},
\newblock \bibinfo{journal}{SIAM Journal on Scientific Computing}
  \bibinfo{volume}{34} (\bibinfo{year}{2012}).
  \DOIprefix\doi{10.1137/120861187}.
%Type = Article
\bibitem[{Bihlo and MacLachlan(2017)}]{bk:Bihlo_numeric[2017]}
\bibinfo{author}{A.~Bihlo}, \bibinfo{author}{S.~MacLachlan},
\newblock \bibinfo{title}{Well-balanced mesh-based and meshless schemes for the
  shallow-water equations},
\newblock \bibinfo{journal}{BIT Numerical Mathematics}  (\bibinfo{year}{2017}).
%Type = Article
\bibitem[{Brecht et~al.(2019)Brecht, Bauer, Bihlo, Gay-Balmaz, and
  MacLachlan}]{bk:Bihlo_numeric[2019]}
\bibinfo{author}{R.~Brecht}, \bibinfo{author}{W.~Bauer},
  \bibinfo{author}{A.~Bihlo}, \bibinfo{author}{F.~Gay-Balmaz},
  \bibinfo{author}{S.~MacLachlan},
\newblock \bibinfo{title}{Variational integrator for the rotating shallow-water
  equations on the sphere},
\newblock \bibinfo{journal}{Quarterly Journal of the Royal Meteorological
  Society} \bibinfo{volume}{145} (\bibinfo{year}{2019}).
  \DOIprefix\doi{10.1002/qj.3477}.
%Type = Article
\bibitem[{{Masum Murshed} et~al.(2019){Masum Murshed}, {Futai}, {Kimura}, and
  {Notsu}}]{bk:MurshedFutai[2019]}
\bibinfo{author}{M.~{Masum Murshed}}, \bibinfo{author}{K.~{Futai}},
  \bibinfo{author}{M.~{Kimura}}, \bibinfo{author}{H.~{Notsu}},
\newblock \bibinfo{title}{{Theoretical and numerical studies for energy
  estimates of the shallow water equations with a transmission boundary
  condition}},
\newblock \bibinfo{journal}{arXiv e-prints}  (\bibinfo{year}{2019})
  \bibinfo{pages}{arXiv:1901.05725}.
  \href{http://arxiv.org/abs/1901.05725}{{\tt arXiv:1901.05725}}.
%Type = Article
\bibitem[{{Khakimzyanov} et~al.(2017){Khakimzyanov}, {Dutykh}, and
  {Gusev}}]{bk:KhakimzyanovIV}
\bibinfo{author}{G.~{Khakimzyanov}}, \bibinfo{author}{D.~{Dutykh}},
  \bibinfo{author}{O.~{Gusev}},
\newblock \bibinfo{title}{{Dispersive shallow water wave modelling. Part IV:
  Numerical simulation on a globally spherical geometry}},
\newblock \bibinfo{journal}{arXiv e-prints}  (\bibinfo{year}{2017})
  \bibinfo{pages}{arXiv:1707.02552}.
  \href{http://arxiv.org/abs/1707.02552}{{\tt arXiv:1707.02552}}.
%Type = Inproceedings
\bibitem[{Dyakonova et~al.(2016)Dyakonova, Khoperskov, and
  Khrapov}]{bk:DyakonovaKhoperskov}
\bibinfo{author}{T.~Dyakonova}, \bibinfo{author}{A.~Khoperskov},
  \bibinfo{author}{S.~Khrapov},
\newblock \bibinfo{title}{Numerical model of shallow water: The use of nvidia
  cuda graphics processors},
\newblock in: \bibinfo{editor}{V.~Voevodin}, \bibinfo{editor}{S.~Sobolev}
  (Eds.), \bibinfo{booktitle}{Supercomputing}, \bibinfo{publisher}{Springer
  International Publishing}, \bibinfo{address}{Cham}, \bibinfo{year}{2016}, pp.
  \bibinfo{pages}{132--145}.
%Type = Article
\bibitem[{Luna et~al.(2013)Luna, Castro~Díaz, and Parés}]{bk:MoralesCastro}
\bibinfo{author}{T.~M. Luna}, \bibinfo{author}{M.~J. Castro~Díaz},
  \bibinfo{author}{C.~Parés},
\newblock \bibinfo{title}{Reliability of first order numerical schemes for
  solving shallow water system over abrupt topography},
\newblock \bibinfo{journal}{Applied Mathematics and Computation}
  \bibinfo{volume}{219} (\bibinfo{year}{2013}) \bibinfo{pages}{9012 -- 9032}.
  \DOIprefix\doi{https://doi.org/10.1016/j.amc.2013.03.033}.
%Type = Article
\bibitem[{Lie(1880)}]{Lie15}
\bibinfo{author}{S.~Lie},
\newblock \bibinfo{title}{Theorie der transformationsgruppen i},
\newblock \bibinfo{journal}{Mathematische Annalen} \bibinfo{volume}{16}
  (\bibinfo{year}{1880}) \bibinfo{pages}{441--528}.
  \DOIprefix\doi{10.1007/BF01446218}.
%Type = Article
\bibitem[{Lie and von Dr.~G.~Scheffers(1891)}]{bk:Lie[1891b]}
\bibinfo{author}{S.~Lie}, \bibinfo{author}{von Dr.~G.~Scheffers},
\newblock \bibinfo{title}{Vorlesungen uber differentialgleichungen mit
  bekannten infinitesimalen transformationen, bearbeitet und herausgegehen},
\newblock \bibinfo{journal}{Mathematische Annalen}  (\bibinfo{year}{1891}).
%Type = Book
\bibitem[{Lie(1896)}]{bk:Lie-Scheffers[1896]}
\bibinfo{author}{S.~Lie}, \bibinfo{title}{Geometrie der
  Ber\"uhrungstransformationen}, \bibinfo{publisher}{B.G. Teubner},
  \bibinfo{address}{Leipzig}, \bibinfo{year}{1896}. \bibinfo{note}{Dargestellt
  von Sophus Lie und Georg Scheffers}.
%Type = Book
\bibitem[{Ovsiannikov(1982)}]{bk:Ovsyannikov[1962]}
\bibinfo{author}{L.~V. Ovsiannikov}, \bibinfo{title}{Group Analysis of
  Differential Equations}, \bibinfo{publisher}{Academic}, \bibinfo{address}{New
  York}, \bibinfo{year}{1982}.
%Type = Book
\bibitem[{Olver(1986)}]{bk:Olver}
\bibinfo{author}{P.~J. Olver}, \bibinfo{title}{Applications of Lie Groups to
  Differential Equations}, \bibinfo{publisher}{Springer}, \bibinfo{address}{New
  York}, \bibinfo{year}{1986}.
%Type = Book
\bibitem[{Ibragimov(1985)}]{bk:Ibragimov1985}
\bibinfo{author}{N.~H. Ibragimov}, \bibinfo{title}{Transformation Groups
  Applied to Mathematical Physics}, \bibinfo{publisher}{Reidel},
  \bibinfo{address}{Boston}, \bibinfo{year}{1985}.
%Type = Book
\bibitem[{Bluman and Kumei(2013)}]{bk:Bluman1989}
\bibinfo{author}{G.~Bluman}, \bibinfo{author}{S.~Kumei},
  \bibinfo{title}{Symmetries and Differential Equations}, Applied Mathematical
  Sciences, \bibinfo{publisher}{Springer New York}, \bibinfo{year}{2013}.
%Type = Book
\bibitem[{Ibragimov(1994)}]{bk:HandbookLie_v1}
\bibinfo{editor}{N.~H. Ibragimov} (Ed.), \bibinfo{title}{{CRC} Handbook of
  {L}ie Group Analysis of Differential Equations}, volume~\bibinfo{volume}{1},
  \bibinfo{publisher}{CRC Press}, \bibinfo{address}{Boca Raton},
  \bibinfo{year}{1994}.
%Type = Book
\bibitem[{Gaeta(1994)}]{bk:Gaeta1994}
\bibinfo{author}{G.~Gaeta}, \bibinfo{title}{Nonlinear Symmetries and Nonlinear
  Equations}, \bibinfo{publisher}{Kluwer}, \bibinfo{address}{Dordrecht},
  \bibinfo{year}{1994}.
%Type = Book
\bibitem[{Ibragimov(1995)}]{bk:HandbookLie_v2}
\bibinfo{editor}{N.~H. Ibragimov} (Ed.), \bibinfo{title}{{CRC} Handbook of
  {L}ie Group Analysis of Differential Equations}, volume~\bibinfo{volume}{2},
  \bibinfo{publisher}{CRC Press}, \bibinfo{address}{Boca Raton},
  \bibinfo{year}{1995}.
%Type = Article
\bibitem[{Levi et~al.(1989)Levi, Nicci, Rogers, and
  Winternitz}]{bk:LeviNicciRogersWint[1989]}
\bibinfo{author}{D.~Levi}, \bibinfo{author}{M.~Nicci},
  \bibinfo{author}{C.~Rogers}, \bibinfo{author}{P.~Winternitz},
\newblock \bibinfo{title}{Group theoretical analysis of a rotating shallow
  liquid in a rigid container},
\newblock \bibinfo{journal}{Journal of Physics A:~Mathematical and General}
  \bibinfo{volume}{22} (\bibinfo{year}{1989}) \bibinfo{pages}{4743--4767}.
  \DOIprefix\doi{10.1088/0305-4470/22/22/007}.
%Type = Article
\bibitem[{Bila et~al.(2006)Bila, Mansfield, and
  Clarkson}]{bk:ClarksonBila[2006]}
\bibinfo{author}{N.~Bila}, \bibinfo{author}{E.~Mansfield},
  \bibinfo{author}{P.~Clarkson},
\newblock \bibinfo{title}{Symmetry group analysis of the shallow water and
  semi-geostrophic equations},
\newblock \bibinfo{journal}{The Quarterly Journal of Mechanics and Applied
  Mathematics} \bibinfo{volume}{59} (\bibinfo{year}{2006}).
  \DOIprefix\doi{10.1093/qjmam/hbi033}.
%Type = Article
\bibitem[{Aksenov and Druzhkov(2020)}]{bk:AksenovDruzkov_classif[2019]}
\bibinfo{author}{A.~V. Aksenov}, \bibinfo{author}{K.~P. Druzhkov},
\newblock \bibinfo{title}{Conservation laws of the equation of one-dimensional
  shallow water over uneven bottom in {L}agrange’s variables},
\newblock \bibinfo{journal}{International Journal of Non-Linear Mechanics}
  \bibinfo{volume}{119} (\bibinfo{year}{2020}) \bibinfo{pages}{103348}.
  \DOIprefix\doi{https://doi.org/10.1016/j.ijnonlinmec.2019.103348}.
%Type = Article
\bibitem[{Kaptsov and Meleshko(2020)}]{bk:KaptsovMeleshko_1D_classf[2018]}
\bibinfo{author}{E.~Kaptsov}, \bibinfo{author}{S.~Meleshko},
\newblock \bibinfo{title}{Analysis of the one-dimensional {E}uler--{L}agrange
  equation of continuum mechanics with a {L}agrangian of a special form},
\newblock \bibinfo{journal}{Applied Mathematical Modelling}
  \bibinfo{volume}{77} (\bibinfo{year}{2020}) \bibinfo{pages}{1497 -- 1511}.
  \DOIprefix\doi{https://doi.org/10.1016/j.apm.2019.09.014}.
%Type = Article
\bibitem[{Siriwat et~al.(2016)Siriwat, Kaewmanee, and
  Meleshko}]{bk:SiriwatKaewmaneeMeleshko2016}
\bibinfo{author}{P.~Siriwat}, \bibinfo{author}{C.~Kaewmanee},
  \bibinfo{author}{S.~V. Meleshko},
\newblock \bibinfo{title}{Symmetries of the hyperbolic shallow water equations
  and the {G}reen-{N}aghdi model in {L}agrangian coordinates},
\newblock \bibinfo{journal}{International Journal of Non-Linear Mechanics}
  \bibinfo{volume}{86} (\bibinfo{year}{2016}) \bibinfo{pages}{185--195}.
%Type = Article
\bibitem[{Szatmari and Bihlo(2014)}]{bk:SzatmariBihlo[2014]}
\bibinfo{author}{S.~Szatmari}, \bibinfo{author}{A.~Bihlo},
\newblock \bibinfo{title}{Symmetry analysis of a system of modified
  shallow-water equations},
\newblock \bibinfo{journal}{Communications in Nonlinear Science and Numerical
  Simulation} \bibinfo{volume}{19} (\bibinfo{year}{2014})
  \bibinfo{pages}{530--537}.
%Type = Article
\bibitem[{Maeda(1985)}]{Maeda1}
\bibinfo{author}{S.~Maeda},
\newblock \bibinfo{title}{Extension of discrete {Noether} theorem},
\newblock \bibinfo{journal}{Math. Japonica} \bibinfo{volume}{26}
  (\bibinfo{year}{1985}) \bibinfo{pages}{85--90}.
%Type = Article
\bibitem[{Maeda(1987)}]{Maeda2}
\bibinfo{author}{S.~Maeda},
\newblock \bibinfo{title}{The similarity method for difference equations},
\newblock \bibinfo{journal}{J. Inst. Math. Appl.} \bibinfo{volume}{38}
  (\bibinfo{year}{1987}) \bibinfo{pages}{129--134}.
%Type = Article
\bibitem[{Dorodnitsyn(1991)}]{Dor_1}
\bibinfo{author}{V.~A. Dorodnitsyn},
\newblock \bibinfo{title}{Transformation groups in net spaces},
\newblock \bibinfo{journal}{Journal of Soviet Mathematics} \bibinfo{volume}{55}
  (\bibinfo{year}{1991}) \bibinfo{pages}{1490--1517}.
  \DOIprefix\doi{10.1007/BF01097535}.
%Type = Article
\bibitem[{Dorodnitsyn(1994)}]{Dor_2}
\bibinfo{author}{V.~A. Dorodnitsyn},
\newblock \bibinfo{title}{Finite difference models entirely inheriting symmetry
  of original differential equations},
\newblock \bibinfo{journal}{International Journal of Modern Physics C}
  \bibinfo{volume}{5} (\bibinfo{year}{1994}).
  \DOIprefix\doi{10.1142/S0129183194000830}.
%Type = Article
\bibitem[{Dorodnitsyn(1993)}]{Dor_3}
\bibinfo{author}{V.~A. Dorodnitsyn},
\newblock \bibinfo{title}{The finite-difference analogy of {Noether}'s
  theorem},
\newblock \bibinfo{journal}{Phys. Dokl.} \bibinfo{volume}{38}
  (\bibinfo{year}{1993}) \bibinfo{pages}{66--68}.
  \DOIprefix\doi{10.1142/S0129183194000830}.
%Type = Article
\bibitem[{Dorodnitsyn et~al.(2004)Dorodnitsyn, Kozlov, and
  Winternitz}]{bk:DorodKozlovWint[2004]}
\bibinfo{author}{V.~A. Dorodnitsyn}, \bibinfo{author}{R.~V. Kozlov},
  \bibinfo{author}{P.~Winternitz},
\newblock \bibinfo{title}{Continuous symmetries of {L}agrangians and exact
  solutions of discrete equations},
\newblock \bibinfo{journal}{Journal of Mathematical Physics}
  \bibinfo{volume}{45} (\bibinfo{year}{2004}) \bibinfo{pages}{336--359}.
  \DOIprefix\doi{10.1063/1.1625418}.
%Type = Article
\bibitem[{Levi and Winternitz(2005)}]{[LW-2]}
\bibinfo{author}{D.~Levi}, \bibinfo{author}{P.~Winternitz},
\newblock \bibinfo{title}{Continuous symmetries of difference equations},
\newblock \bibinfo{journal}{Journal of Physics A: Mathematical and General}
  \bibinfo{volume}{39} (\bibinfo{year}{2005}) \bibinfo{pages}{R1--R63}.
  \DOIprefix\doi{10.1088/0305-4470/39/2/r01}.
%Type = Article
\bibitem[{Dorodnitsyn et~al.(2000)Dorodnitsyn, Kozlov, and
  Winternitz}]{bk:DorodKozlovWinternitz[2000]}
\bibinfo{author}{V.~A. Dorodnitsyn}, \bibinfo{author}{R.~V. Kozlov},
  \bibinfo{author}{P.~Winternitz},
\newblock \bibinfo{title}{Lie group classification of second-order ordinary
  difference equations},
\newblock \bibinfo{journal}{Journal of Mathematical Physics}
  \bibinfo{volume}{41} (\bibinfo{year}{2000}) \bibinfo{pages}{480--504}.
  \DOIprefix\doi{10.1063/1.533142}.
%Type = Article
\bibitem[{Quispel and Sahadevan(1993)}]{Quisp}
\bibinfo{author}{G.~Quispel}, \bibinfo{author}{R.~Sahadevan},
\newblock \bibinfo{title}{Lie symmetries and the integration of difference
  equations},
\newblock \bibinfo{journal}{Physics Letters A} \bibinfo{volume}{184}
  (\bibinfo{year}{1993}) \bibinfo{pages}{64 -- 70}.
  \DOIprefix\doi{https://doi.org/10.1016/0375-9601(93)90347-3}.
%Type = Article
\bibitem[{Winternitz(2011)}]{[LW-3]}
\bibinfo{author}{P.~Winternitz},
\newblock \bibinfo{title}{Symmetry preserving discretization of differential
  equations and {L}ie point symmetries of differential-difference equations}
  (\bibinfo{year}{2011}).
%Type = Book
\bibitem[{Dorodnitsyn(2011)}]{bk:Dorodnitsyn[2011]}
\bibinfo{author}{V.~A. Dorodnitsyn}, \bibinfo{title}{Applications of Lie Groups
  to Difference Equations}, \bibinfo{publisher}{CRC Press},
  \bibinfo{address}{Boca Raton}, \bibinfo{year}{2011}.
%Type = Article
\bibitem[{Floreanini and Vinet(1995)}]{Vinet}
\bibinfo{author}{R.~Floreanini}, \bibinfo{author}{L.~Vinet},
\newblock \bibinfo{title}{Lie symmetries of finite‐difference equations},
\newblock \bibinfo{journal}{Journal of Mathematical Physics}
  \bibinfo{volume}{36} (\bibinfo{year}{1995}) \bibinfo{pages}{7024--7042}.
  \DOIprefix\doi{10.1063/1.531205}.
%Type = Book
\bibitem[{Hydon(2014)}]{bk:Hydon_book[2014]}
\bibinfo{author}{P.~E. Hydon}, \bibinfo{title}{Difference Equations by
  Differential Equation Methods}, Cambridge Monographs on Applied and
  Computational Mathematics, \bibinfo{publisher}{Cambridge University Press},
  \bibinfo{year}{2014}. \DOIprefix\doi{10.1017/CBO9781139016988}.
%Type = Article
\bibitem[{Dorodnitsyn et~al.(2015)Dorodnitsyn, Kaptsov, Kozlov, and
  Winternitz}]{bk:DorodKozlovWintKaptsov[2015]}
\bibinfo{author}{V.~A. Dorodnitsyn}, \bibinfo{author}{E.~I. Kaptsov},
  \bibinfo{author}{R.~V. Kozlov}, \bibinfo{author}{P.~Winternitz},
\newblock \bibinfo{title}{The adjoint equation method for constructing first
  integrals of difference equations},
\newblock \bibinfo{journal}{Journal of Physics A: Mathematical and Theoretical}
  \bibinfo{volume}{48} (\bibinfo{year}{2015}) \bibinfo{pages}{055202}.
  \DOIprefix\doi{10.1088/1751-8113/48/5/055202}.
%Type = Article
\bibitem[{Budd et~al.(1996)Budd, Weizhang, and Russell}]{bk:BuddHuang[1996]}
\bibinfo{author}{C.~J. Budd}, \bibinfo{author}{H.~Weizhang},
  \bibinfo{author}{R.~D. Russell},
\newblock \bibinfo{title}{Moving mesh methods for problems with blow-up},
\newblock \bibinfo{journal}{SIAM Journal on Scientific Computing}
  \bibinfo{volume}{17} (\bibinfo{year}{1996}) \bibinfo{pages}{305--327}.
  \DOIprefix\doi{10.1137/S1064827594272025}.
%Type = Book
\bibitem[{Huang and Russel(2010)}]{[Huang]}
\bibinfo{author}{W.~Huang}, \bibinfo{author}{R.~D. Russel},
  \bibinfo{title}{Adaptive Moving Mesh Methods}, \bibinfo{publisher}{Springer},
  \bibinfo{address}{New York}, \bibinfo{year}{2010}.
%Type = Article
\bibitem[{Arakawa and Lamb(1981)}]{bk:ArakawaLamb[1981]}
\bibinfo{author}{A.~Arakawa}, \bibinfo{author}{V.~R. Lamb},
\newblock \bibinfo{title}{A potential enstrophy and energy conserving scheme
  for the shallow water equations},
\newblock \bibinfo{journal}{Monthly Weather Review} \bibinfo{volume}{109}
  (\bibinfo{year}{1981}) \bibinfo{pages}{18--36}.
%Type = Book
\bibitem[{Samarskii and Popov(1980)}]{bk:SamarskyPopov_book[1992]}
\bibinfo{author}{A.~Samarskii}, \bibinfo{author}{Y.~P. Popov},
  \bibinfo{title}{Difference methods for solving problems of gas dynamics},
  \bibinfo{publisher}{Nauka}, \bibinfo{address}{Moscow}, \bibinfo{year}{1980}.
  \bibinfo{note}{In Russian}.
%Type = Book
\bibitem[{Rojdestvenskiy and Yanenko(1968)}]{bk:YanenkRojd[1968]}
\bibinfo{author}{B.~L. Rojdestvenskiy}, \bibinfo{author}{N.~N. Yanenko},
  \bibinfo{title}{Systems of quasilinear equations and their applications to
  gas dynamics}, \bibinfo{publisher}{Nauka}, \bibinfo{address}{Moscow},
  \bibinfo{year}{1968}. \bibinfo{note}{In Russian}.
%Type = Article
\bibitem[{Popov and Samarskiy(1969)}]{bk:SamarskyPopov[1969]}
\bibinfo{author}{Y.~P. Popov}, \bibinfo{author}{A.~A. Samarskiy},
\newblock \bibinfo{title}{Completely conservative difference schemes},
\newblock \bibinfo{journal}{Zh. Vychisl. Mat. Mat. Fiz.} \bibinfo{volume}{9}
  (\bibinfo{year}{1969}) \bibinfo{pages}{953--958}.
%Type = Article
\bibitem[{Koldoba et~al.(1987)Koldoba, Poveschenko, and Popov}]{KolPovPop87}
\bibinfo{author}{A.~V. Koldoba}, \bibinfo{author}{Y.~A. Poveschenko},
  \bibinfo{author}{Y.~P. Popov},
\newblock \bibinfo{title}{Two-layer completely conservative difference schemes
  for the equations of gas dynamics in {E}uler variables},
\newblock \bibinfo{journal}{U.S.S.R. Comput. Math. Math. Phys.}
  \bibinfo{volume}{27} (\bibinfo{year}{1987}) \bibinfo{pages}{91–95}.
%Type = Article
\bibitem[{Poveshchenko et~al.(2019)Poveshchenko, Ladokina, Podryga, Rahimly,
  and Sharova}]{bk:Poveschenko[2019]}
\bibinfo{author}{Y.~A. Poveshchenko}, \bibinfo{author}{M.~E. Ladokina},
  \bibinfo{author}{V.~O. Podryga}, \bibinfo{author}{O.~R. Rahimly},
  \bibinfo{author}{Y.~S. Sharova},
\newblock \bibinfo{title}{On a two-layer completely conservative difference
  scheme of gas dynamics in {E}ulerian variables with adaptive regularization
  of solution},
\newblock \bibinfo{journal}{Keldysh Institute preprints}
  (\bibinfo{year}{2019}).
%Type = Article
\bibitem[{Popov et~al.(2017)Popov, Poveschenko, Polyakov, and
  Rahimli}]{bk:PopovPoveschenkoPolyakov[2017]}
\bibinfo{author}{I.~Popov}, \bibinfo{author}{Y.~Poveschenko},
  \bibinfo{author}{S.~Polyakov}, \bibinfo{author}{P.~Rahimli},
\newblock \bibinfo{title}{One approach to constructing a conservative
  difference scheme for the two-phase filtration problem},
\newblock \bibinfo{journal}{Keldysh Institute preprints}
  (\bibinfo{year}{2017}).
%Type = Article
\bibitem[{Dorodnitsyn et~al.(2019)Dorodnitsyn, Kozlov, and
  Meleshko}]{DORODNITSYN2019201}
\bibinfo{author}{V.~A. Dorodnitsyn}, \bibinfo{author}{R.~V. Kozlov},
  \bibinfo{author}{S.~V. Meleshko},
\newblock \bibinfo{title}{One-dimensional gas dynamics equations of a
  polytropic gas in {L}agrangian coordinates: Symmetry classification,
  conservation laws, difference schemes},
\newblock \bibinfo{journal}{Communications in Nonlinear Science and Numerical
  Simulation} \bibinfo{volume}{74} (\bibinfo{year}{2019}) \bibinfo{pages}{201
  -- 218}. \DOIprefix\doi{https://doi.org/10.1016/j.cnsns.2019.03.009}.
%Type = Article
\bibitem[{{Kozlov}(2019)}]{KOZLOV2019}
\bibinfo{author}{R.~{Kozlov}},
\newblock \bibinfo{title}{{Conservative difference schemes for one-dimensional
  flows of polytropic gas}},
\newblock \bibinfo{journal}{Communications in Nonlinear Science and Numerical
  Simulations} \bibinfo{volume}{78} (\bibinfo{year}{2019})
  \bibinfo{pages}{104864}. \DOIprefix\doi{10.1016/j.cnsns.2019.104864}.
%Type = Article
\bibitem[{Korobitsyn(1989)}]{bk:Korobitsyn_scheme[1989]}
\bibinfo{author}{V.~A. Korobitsyn},
\newblock \bibinfo{title}{Thermodynamically matched difference schemes},
\newblock \bibinfo{journal}{U.S.S.R. Comput. Math. Math. Phys}
  \bibinfo{volume}{29} (\bibinfo{year}{1989}) \bibinfo{pages}{71--79}.
%Type = Article
\bibitem[{Chirkunov and Pikmullina(2014)}]{bk:ChirkunovPikmullina[2014]}
\bibinfo{author}{Y.~Chirkunov}, \bibinfo{author}{E.~O. Pikmullina},
\newblock \bibinfo{title}{Symmetry properties and solutions of shallow water
  equations},
\newblock \bibinfo{journal}{Universal Journal of Applied Mathematics}
  \bibinfo{volume}{2} (\bibinfo{year}{2014}) \bibinfo{pages}{10--23}.
%Type = Book
\bibitem[{Ovsiannikov(2003)}]{Ovsiannikov[2003]}
\bibinfo{author}{L.~V. Ovsiannikov}, \bibinfo{title}{Lectures on the gas
  dynamics equations}, \bibinfo{publisher}{Institute of computer studies},
  \bibinfo{address}{Moscow, Izhevsk}, \bibinfo{year}{2003}.
  \bibinfo{note}{Second edition}.
%Type = Book
\bibitem[{Bluman and Anco(2002)}]{bk:BlumanAnco}
\bibinfo{author}{G.~W. Bluman}, \bibinfo{author}{S.~C. Anco},
  \bibinfo{title}{Symmetry and Integration Methods for Differential Equations},
  \bibinfo{publisher}{Springer}, \bibinfo{address}{New York},
  \bibinfo{year}{2002}.
%Type = Book
\bibitem[{Bluman et~al.(2010)Bluman, Cheviakov, and
  Anco}]{bk:BlumanCheviakovAnco}
\bibinfo{author}{G.~W. Bluman}, \bibinfo{author}{A.~F. Cheviakov},
  \bibinfo{author}{S.~C. Anco}, \bibinfo{title}{Applications of Symmetry
  Methods to Partial Differential Equations}, \bibinfo{publisher}{Springer},
  \bibinfo{address}{New York}, \bibinfo{year}{2010}. \bibinfo{note}{Applied
  Mathematical Sciences, Vol.168}.
%Type = Article
\bibitem[{Dorodnitsyn(2006)}]{bk:DorodLinearization}
\bibinfo{author}{V.~A. Dorodnitsyn},
\newblock \bibinfo{title}{On the linearization of second-order differential and
  difference equations},
\newblock \bibinfo{journal}{Symmetry, Integrability and Geometry: Methods and
  Applications} \bibinfo{volume}{2} (\bibinfo{year}{2006}).
  \DOIprefix\doi{10.3842/SIGMA.2006.065}.
%Type = Article
\bibitem[{Samarskiy and Popov(1970)}]{bk:SamarskyPopov[1970]}
\bibinfo{author}{A.~A. Samarskiy}, \bibinfo{author}{Y.~P. Popov},
\newblock \bibinfo{title}{Completely conservative difference schemes for the
  equations of magneto-hydrodynamics},
\newblock \bibinfo{journal}{U.S.S.R. Comput. Math. Math. Phys.}
  \bibinfo{volume}{10} (\bibinfo{year}{1970}) \bibinfo{pages}{233--243}.
%Type = Article
\bibitem[{Kaptsov(2019)}]{bk:EK_preprint[2019]}
\bibinfo{author}{E.~I. Kaptsov},
\newblock \bibinfo{title}{Numerical implementation of an invariant scheme for
  one-dimensional shallow water equations in {L}agrangian coordinates},
\newblock \bibinfo{journal}{Keldysh Institute preprints}
  (\bibinfo{year}{2019}). \bibinfo{note}{In Russian}.

\end{thebibliography}
%\inputencoding{cp866}

\end{document}